\documentclass[12pt,preprint]{aastex}

\usepackage{subfigure}
\usepackage{booktabs}
\usepackage{threeparttable}
\usepackage{filecontents}
\usepackage{array}
\usepackage{float}
\DeclareGraphicsExtensions{.eps}
\usepackage[papersize={8.5in,11in}, top=1.5cm, bottom=1.5cm, left=1.5cm, right=1.5cm]{geometry}

\shorttitle{Superflares of Kepler-96 and their biological impact onto the planetary system}
\shortauthors{Estrela and Valio}

\begin{document}


\title{Superflare UV impact on Kepler-96 system: a glimpse of habitability when the ozone layer first formed on Earth}


\author{Raissa Estrela and Adriana Valio}

\affil{Center for Radio Astronomy and Astrophysics (CRAAM),
Mackenzie Presbyterian University, Rua da Consola\c c\~o 896, S\~ao Paulo,
SP 01302-907, Brazil}
\email{rlf.estrela@gmail.com}
\email{avalio@craam.mackenzie.br}






\begin{abstract}
Kepler-96 is an active solar-type star harbouring a Super-Earth planet in close orbit. Its age of 2.3 Gyr is the same as the Sun when there was a considerable increase of oxygen in Earth's atmosphere due to micro-organisms living in the ocean. We present the analysis of superflares seen on the transit lightcurves of Kepler-96b. The model used here simulates the planetary transit in a flaring star. By fitting the observational data with this model, it is possible to infer the physical properties of the flares, such as their duration and the energy released. We found 3 flares within the energy range of superflares, where the biggest superflare observed was found to have an energy of 1.81$\times$10$^{35}$ ergs. The goal is to analyse the biological impact of these superflares on a hypothetical Earth in the habitable zone of Kepler-96 assuming this planet has protection via different scenarios: an Archean  and Present-day atmospheres. Also, we compute the attenuation of the flare UV radiation through an Archean ocean. The conclusion is that considering the increase in the UV flux by the strongest superflare emission, {\it E. Coli} and {\it D. Radiodurans} could survive on the surface of the planet only if there was an ozone layer present on the planet atmosphere. However, they could escape from the hazardous UV effects at a depth of 28m and 12m below the ocean surface, respectively. For smaller superflares contribution, {\it D. Radiodurans} could survive in the surface even in an Archean atmosphere with no ozone.
\end{abstract}


\keywords{superflares, magnetic activity, archean, UV radiation, planetary atmosphere, habitability}


\section*{Introduction}

Observations by the Kepler telescope have led to the discovery of stars that are similar to our Sun. These stars can be very young, only a few million years old, or  have an intermediate age of a few giga-years. Kepler data showed that these solar analogues can be very active and have flares much more powerful than those on the present Sun. The occurrence of  flares and the energy released will depend on the magnetic field strength of the star. The chromospheric activity of a star is related to its rotation rate and consequently to its age. Young rapidly rotating stars show high level of activity and can produce large starspots and energetic flares. On the other hand, slow rotating stars, like our Sun, have fewer and smaller spots and lower flare activity. It is also expected that younger stars frequently produce  superflares \citep{maehara13}. However, contrary to expectations,  these phenomena were  observed in stars with temperature and rotation similar to that of our Sun \citep{maehara15}. 

The vast amount of superflares found in young Kepler solar-like stars have led researchers to wonder about the  occurrence of such phenomena on the young  Sun. Moreover, the impact of these superflares on life in the primitive Earth was also inquired. \cite{airapetian16} suggested that superflares may have had a positive influence for the beginning of life on Earth. They showed that superflares allied with energetic particles can change the chemistry of the early Earth's atmosphere producing greenhouse gases responsible for warming Earth at a time when the Sun was not luminous enough. On the other hand, \cite{lingam17}  argue that powerful superflares can lead to extinction events, and their periodicity coincides with terrestrial fossil extinction record.  Indeed, superflares can release a significant amount of EUV and UV radiation, which could cause potential effects on  planetary atmospheres or biological processes affecting the origin and evolution of life \citep{segura10}.

The star analysed here, Kepler-96, is a solar analogue with an age of 2.3 Gyr. This is an intermediate age for a star, however Kepler-96 is still very active and shows a clear rotational modulation in its lightcurve. In particular, the lightcurve of the star is marked by several flares, which confirm the high level activity of the star. Kepler-96 has a super-Earth in its orbit that transits this star once every orbital period of 16.23 days. The transits of this planet also show signatures of flares, and in particular an extremely significant flare peak is visible in the middle of the 48th transit (966.70 BJD - 2.454.833 days)).

Interestingly, the star has an age that corresponds to the start of the Proterozoic Era on Earth. This period coincides with the beginning of oxygen level increase in Earth's atmosphere due to the photosynthesis of cyanobacteria living in the ocean. The accumulation of the oxygen allowed the formation of the ozone layer, which is responsible for absorbing most of the solar ultraviolet radiation arriving at Earth. In particular those radiations that are more threatening for life like UVC (100-280 nm) and UVB (280-315 nm), with only UVA (315-400nm) reaching the surface. Despite the fact that planets can follow different evolutionary paths, here we supose that a planet orbiting Kepler-96 could have had enough time to develop multicellular or single-celled lifeforms (such as cyanobacterias on Earth) that helps the prodution of an ozone layer. This would be an essential factor to protect the planet from the flare hazards of its host star since a significant amount of EUV and UV radiation can be released during these events. In this context, this star can be used as a proxy to understand both: (1) the primitive Earth environment, assuming  the Sun  also produced superflares at that time, and  (2) a planet in the habitable zone with Archean conditions, considering that this planet had already enough time to evolve. 


Here, we analyse the flares of Kepler-96 using the transit model developed by \cite{sil03}. This model allows the addition of stellar features such as spots or flares in the simulation. We propose two scenarios for the detection of a flare during a planetary transit: a flare occulted by the planet during transit and a flare outside the transit band. Properties of the flares, such as duration, size, amplitude, and position will be intrinsically studied with this model. From these parameters it is possible to estimate the power released by the flare. Previously, the occurrence of a stellar flare during the transit of a planet was reported only by \cite{bentley09}, however its characteristics were not fully described.

The contribution of the flares to the stellar UV flux is also inferred. Next, we study how different atmospheres of the primitive and present-day Earth would attenuate the increased UV flux arriving at the surface of a planet in the habitable zone (HZ, at 1AU) of Kepler-96. Depending on its intensity, the UV flux could turn the planet into a hostile environment that would not allow the development of life on its surface. In addition, we also analyse the possibility of life survival in the ocean, by determining the minimum ocean depth where extremophile lifeforms are not damaged by the UV radiation. The effects in Kepler-96b are not analysed here because due to its proximity to the host star this planet must be under very extreme conditions and should not be habitable. For example, assuming a Bond albedo of 0.3, the equilibrium temperature of Kepler-96b should be around 3700 K. Furthermore, superflares allied with CMEs could be stripping away the atmosphere of this planet.

This paper is organised as follows. Section \ref{sec:sec2} provides the details of the Kepler mission observations of Kepler-96. Section \ref{sec:sec3} explains the model adopted in this work for simulating transits and shows the different situations of flare detection within planetary transits. The results for the flares parameters using the transit model of \cite{sil03} are shown in Section \ref{sec:32}. The biological impact of the flaring UV flux for life either on the surface or in the ocean of the hypothetical planet in the HZ is evaluated in Section~\ref{bio}. Finally, Section \ref{sec:sec5} concludes the paper with the main results and discussions.

\section{Observations} 
\label{sec:sec2}

The Kepler mission collected data for thousands of stars located in the constellation of Cygnus, Lyra, and Draco. The duration of the mission was scheduled for 3.5 years but was extended to 2016 with the K2 mission (see \cite{howell14}). The telescope has an excellent photometry and 115$\rm deg^{2}$ field of view. Up to now, a total of 2337 planets detected by the Kepler telescope were confirmed and there are a total of 4496 candidates.

The target stars of the Kepler mission were observed in long (one data point each 29.4 minutes) and short cadence ($\sim$1 minute time resolution). These two types of lightcurves can be obtained in the MAST database\footnote{http://archive.stsci.edu/} (\textit{Mikulski Archive for Space Telescopes}). Here, we  analyse only the short cadence lightcurves because we are interested in short-timescale variations within planetary transits over the whole period of operation of the satellite.

Kepler-96, the star analysed in this work, was observed by the Kepler telescope during a total of 1364 days. Kepler-96 is an active star exhibiting a lightcurve  marked by rotational modulation and flare activity. This star has approximately the same mass, $M_* = 1.00 \pm 0.06 M_\odot$, and size as the Sun with radius $R_{*} = 1.02 \pm 0.09 R_\odot$, but with no determined spectral type. It has an estimated age of 2.34 Gyr and is accompanied by a super-Earth planet in close orbit. Kepler-96b has $0.027 M_{\rm Jup}$ and an orbital period of 16.23 days \citep{marcy14}. The main parameters of the star are listed on Table~\ref{table:tab1}. During the time of observation of the $Kepler$ mission, a total of 86 planetary transits were detected, and in particular the 48th transit showed a very intense flare.

All the transit lightcurves were fitted using the model from \cite{sil03} and the parameters of the planet and orbit (radius, semimajor axis, and inclination angle) were derived by applying a least chi-square minimisation routine (AMOEBA) to the data. The value found for the radius of the planet is 2$\%$ larger than that reported by \cite{marcy14}, due to the activity of the star, such as spots, which makes the transit seem shallower. The same phenomena was observed in the transits of CoroT-2 and Kepler-17 \citep{sil10, valio16}. The parameters of the planet are given in Table \ref{table:tab2}.

\begin{figure*}[!ht]
  \centering
\includegraphics[scale=0.25]{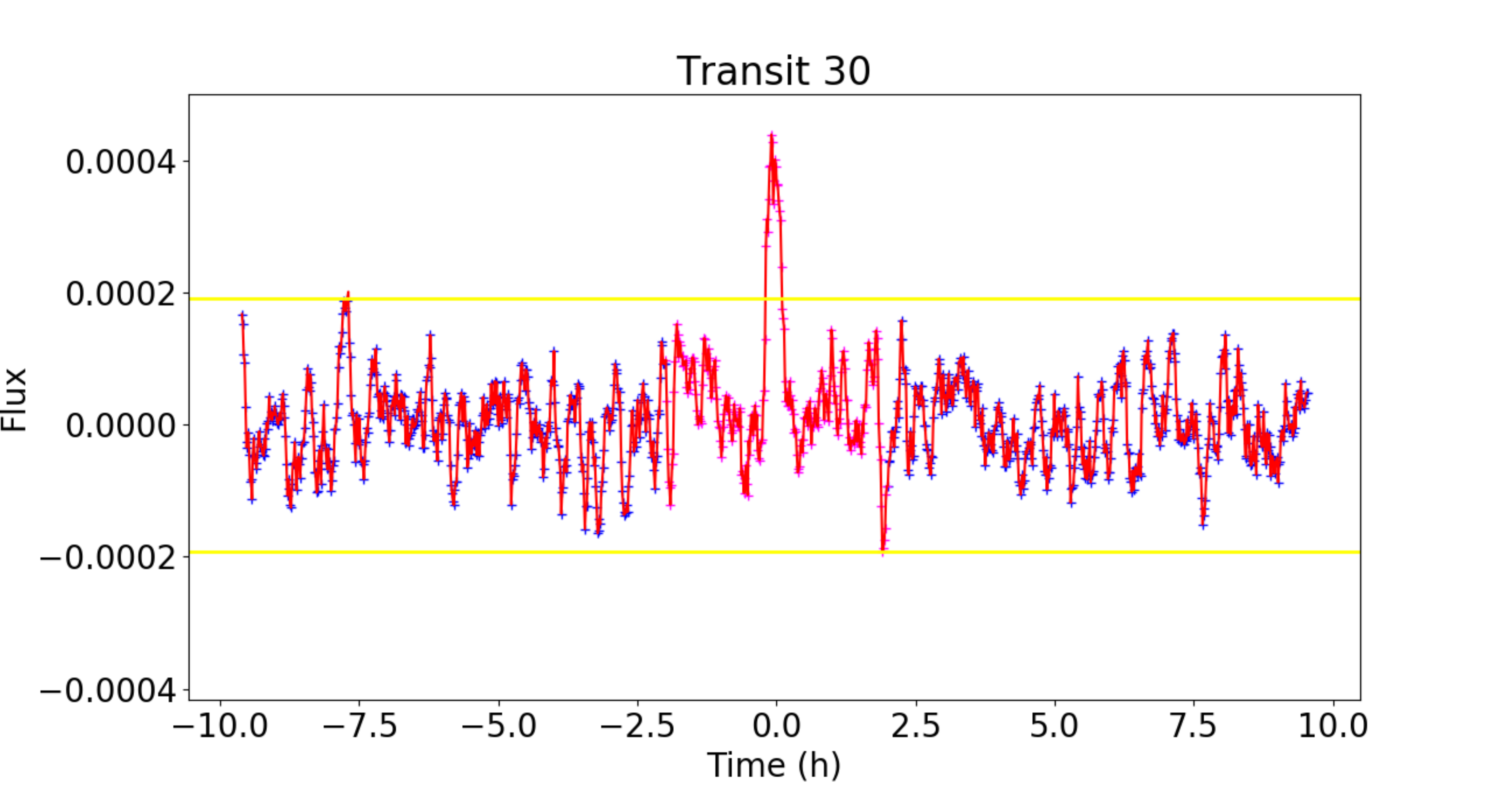} \\
\includegraphics[scale=0.25]{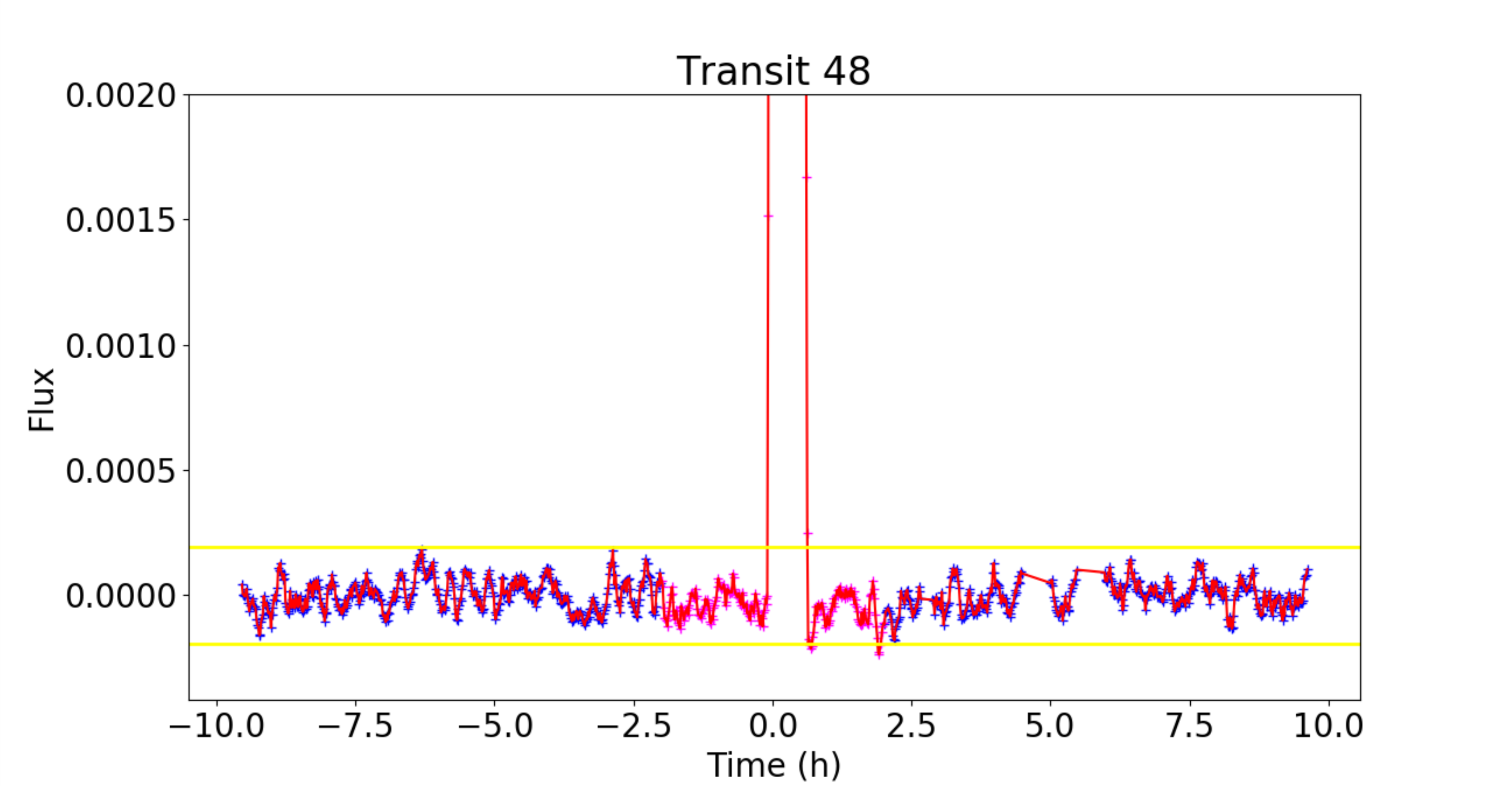} \\
\includegraphics[scale=0.25]{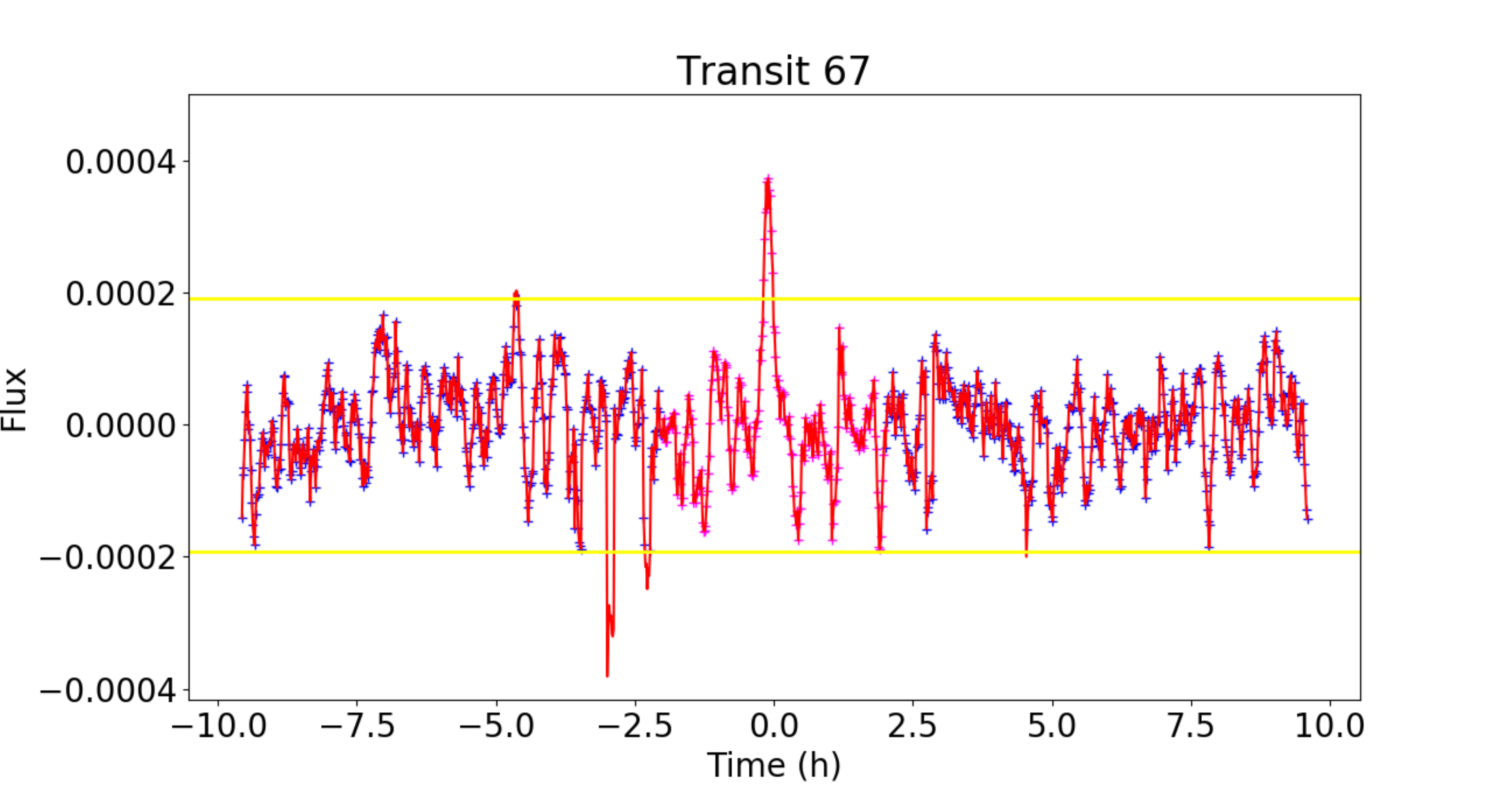} \\
  \caption[Folded]{Noise correction in the transit lightcurves by application of a sigma-clip procedure. Points above or below $\pm$ 10 CDPP (yellow horizontal line) are considered as noise and removed, only for points outside the planetary transits (marked by a blue cross). Points within the transits are marked by a pink cross.}
\label{fig:figura1_1}
\end{figure*}

To remove random noise in the transits lightcurves, we adopted the sigma clipping technique. We set a threshold of 10 times the  Combined Differential Photometric Precision (CDPP) of Kepler-96, which is an estimate of the noise in Kepler data \citep{christiansen12}, and datapoints above or below the threshold were considered as outliers and rejected. Figure \ref{fig:figura1_1} shows the application of the sigma clipping technique to all the transits lightcurves analysed in this work. The rejected values that exceeded the threshold (dotted horizontal line in \ref{fig:figura1_1}) are not marked by a cross symbol. This technique is not applied to those points that fall inside the transit (marked by a red cross in Fig. \ref{fig:figura1_1}) so as not to remove the flare signal. Moreover, only the transit lightcurve between ingress and egress times of the planet were analysed, this corresponds to $\pm 70^{\circ}$ longitude projected on the stellar surface.


\section{Modelling flares in transit}
\label{sec:sec3}

\subsection{Method}\label{sec:sec31}

In this work, we characterise the flares present in the transit lightcurves of Kepler-96 using the transit model proposed by \cite{sil03}. This model simulates a star with quadratic limb darkening as a 2D image and the planet is assumed to be a dark disk with radius $R_{p}/R_{star}$, where $R_{p}$ is the radius of the planet and $R_{star}$ is the radius of the host star. The simulations are performed considering a circular orbit (null eccentricity). For a given time interval,  the position of the planet in its orbit is calculated according to the semi-major axis and inclination angle. The simulated lightcurve is obtained by the sum of the intensity of all the pixels in the image (star plus planet) and normalised.

In addition, this model allows the insertion of features in the stellar disc, such as spots, faculae, and flares. Applications of this model to characterise starspots and analysis of differential rotation in active stars have already been performed by \cite{sil08,sil10,netto16,valio16}. Also, this model was used to study magnetic cycles in the stars Kepler-17 and Kepler-63 \citep{estrela16}. Here, we will add a flare to the stellar surface. 

Therefore, we propose for the first time a simulation of a flare being occulted by a planet transiting its host star. In this model, we consider a flare that has a duration equal to or higher than the transit  duration ($\sim$ 2 hours). Large flares on the Sun are known to have this duration. Moreover, \cite{maehara15} analysed long and short cadence lightcurves from Kepler data and found flares with a bolometric energy range of 2 $\times 10^{32}$ up to 8 $\times 10^{35}$ ergs, and amplitudes ranging from 0.06$\%$ up to $8\%$ of the star luminosity, and a duration interval varying from 5 to 120 min. 

\begin{figure}[!ht]
  \centering
\includegraphics[scale=0.40]{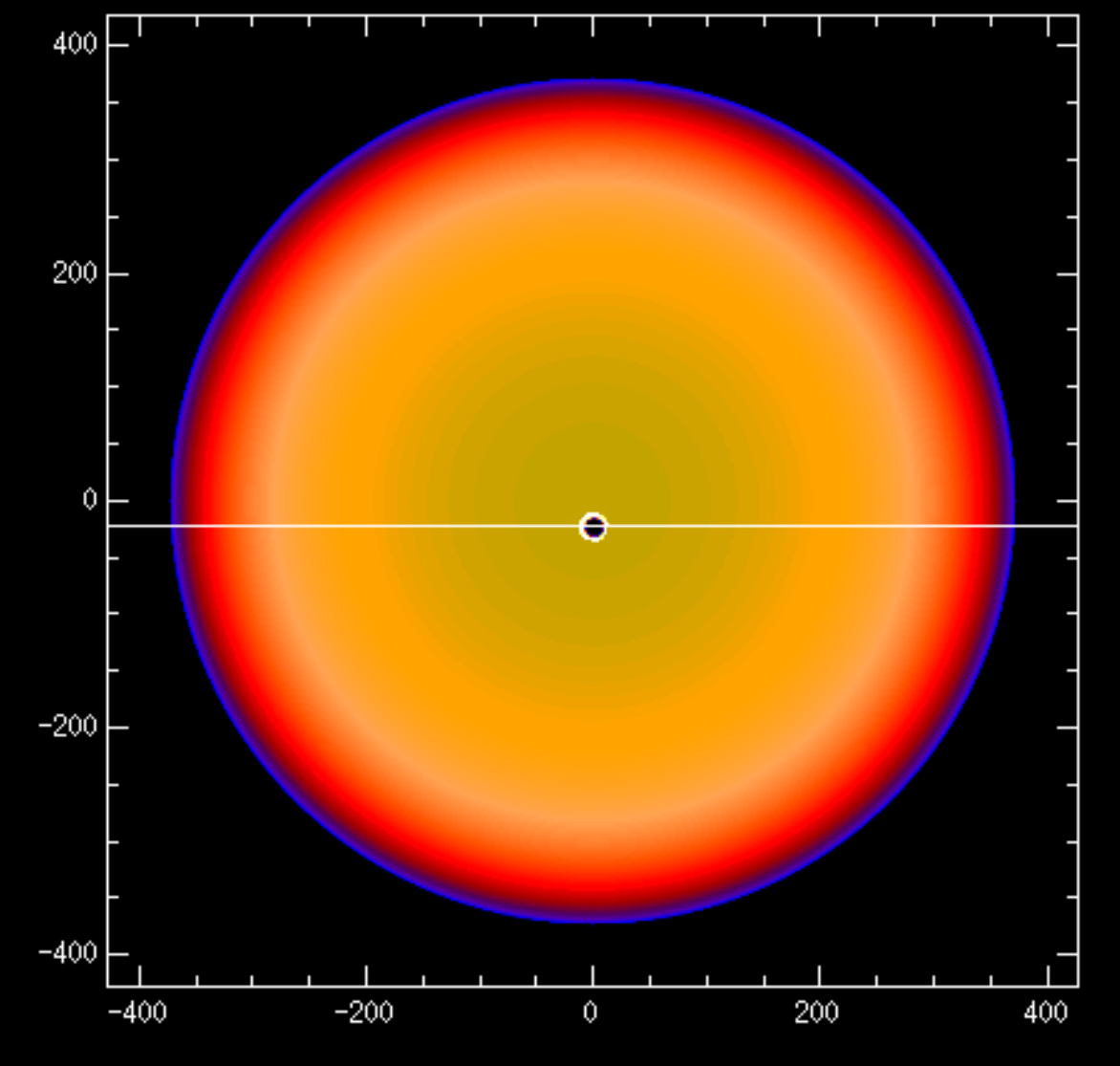} \\
\includegraphics[scale=0.35]{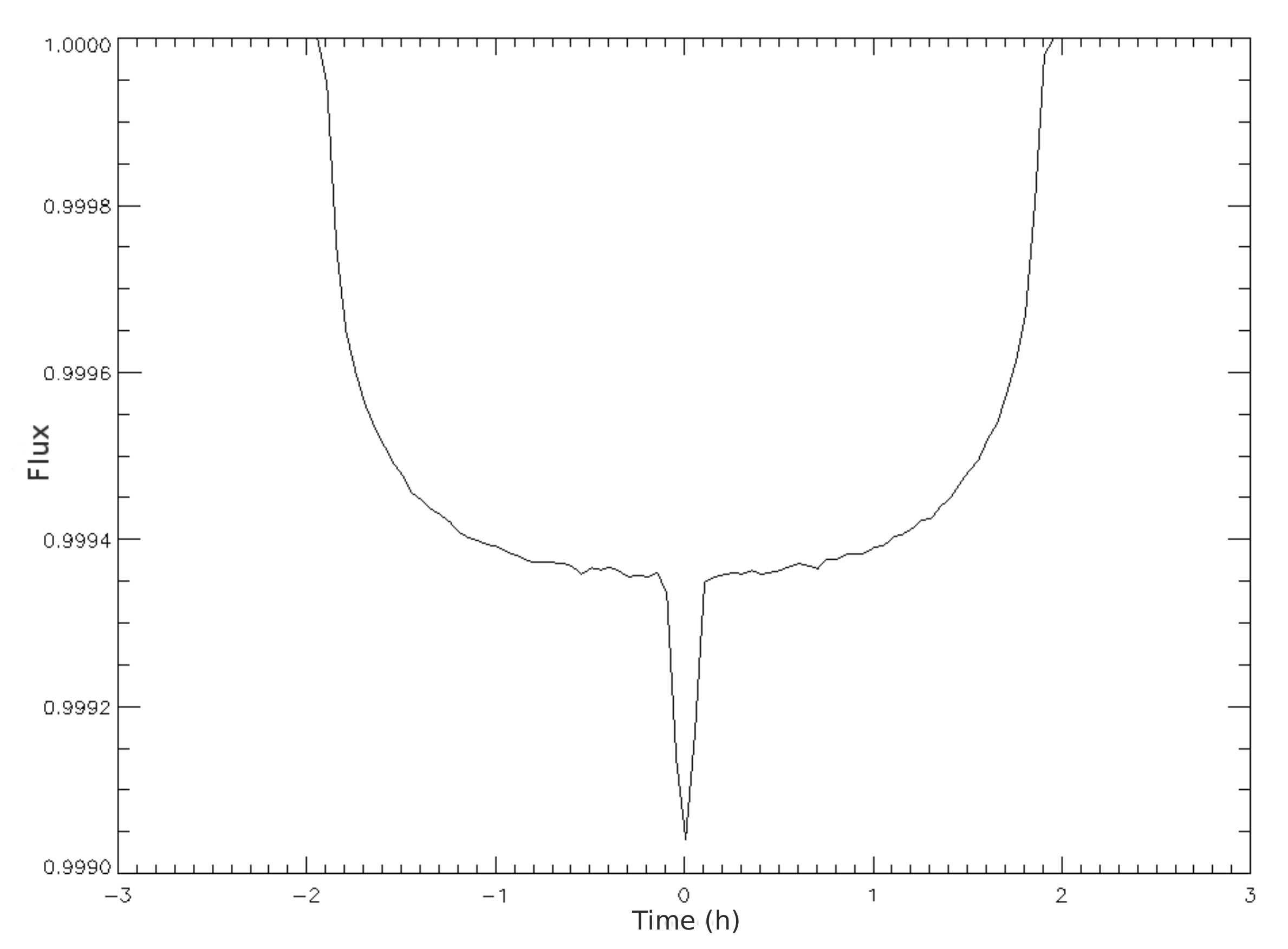}
  \caption[Folded]{{\it Top:} Synthesized star with a flare (white circle) eclipsed by the planet (dark circle). {\it Bottom:} 	}
\label{fig:figura1_2}
\end{figure}

The transit of the planet causes a decrease in the stellar luminosity when this planet passes in front of the stellar flaring region  located in the disk of the star, producing a negative peak in the transit lightcurve. This occurs because the flare is an extremely bright phenomena compared to the rest of the stellar surface, similar to white light flares on the Sun. The effect produced by the presence of an occulted flare is illustrated in Fig. \ref{fig:figura1_2}. As showed by \cite{sil08,sil10}, this signature is exactly the opposite to what happens when a planet occults a starspot (dark and cooler region). In the latter case, one expects a slight increase in the luminosity (a ``bump'') in the transit lightcurve. This phenomenum occurs because the occulted spot is a region darker than the photosphere.

\begin{figure}[!ht]
  \centering
\includegraphics[scale=0.40]{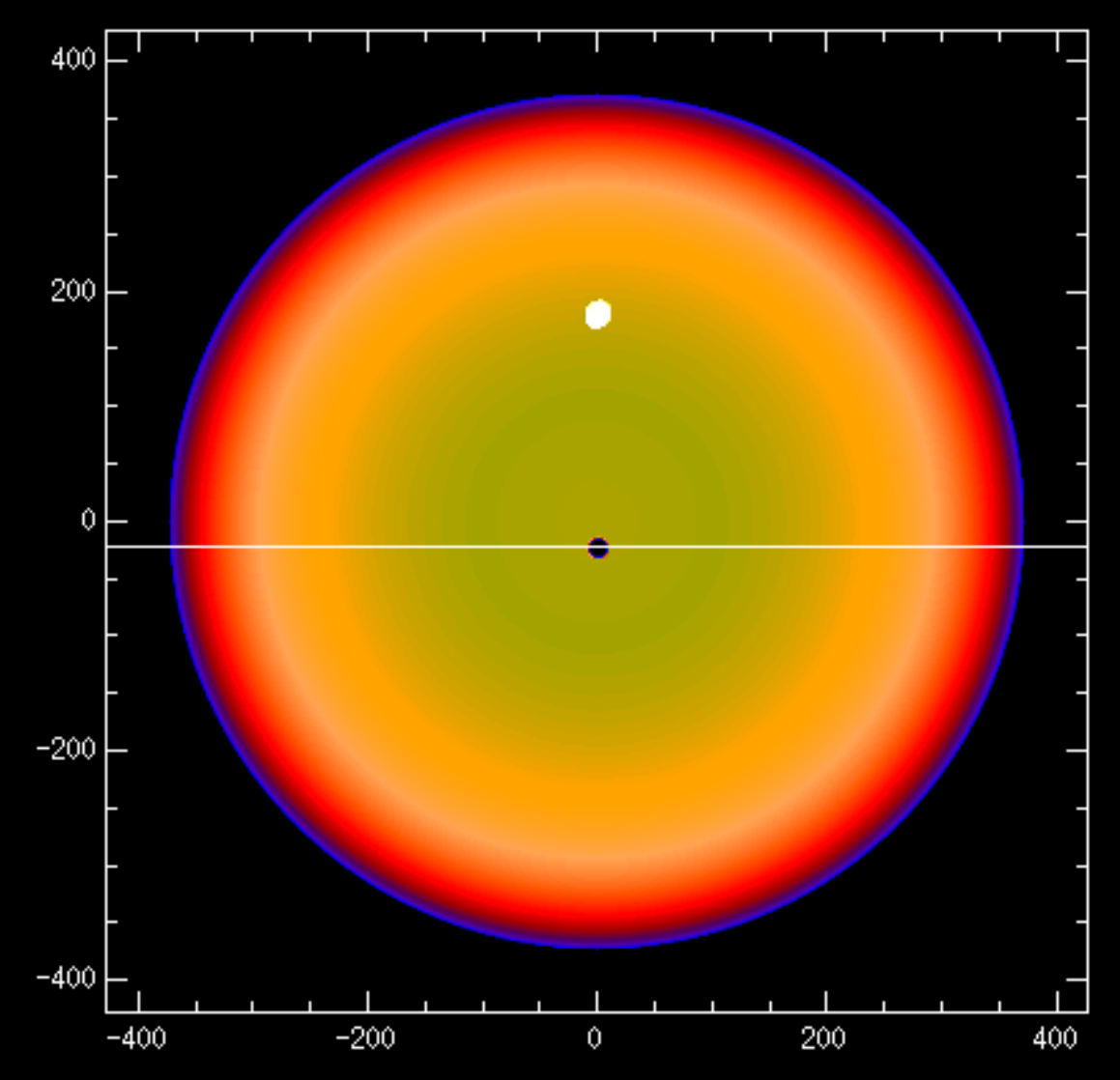} \\
\includegraphics[scale=0.35]{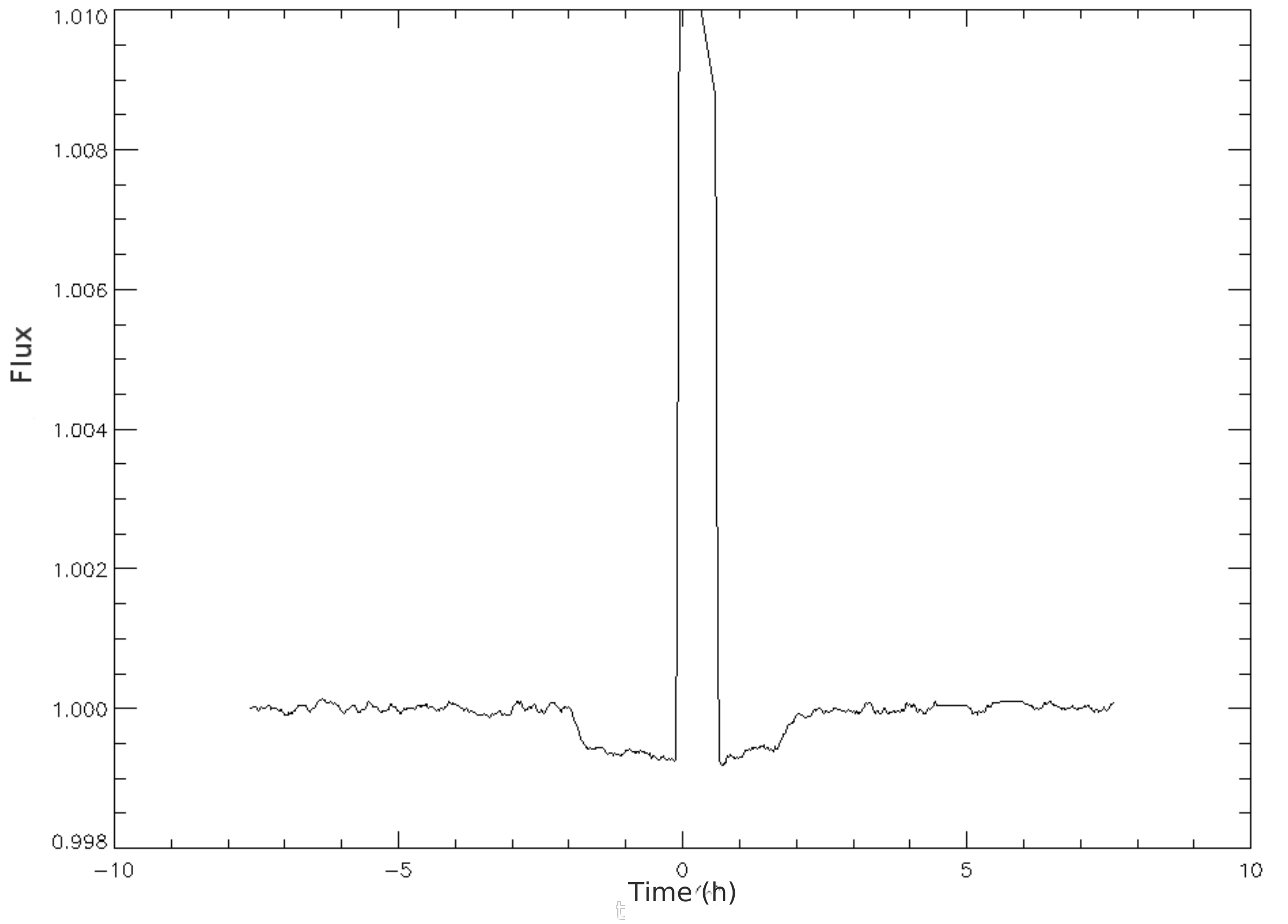}
  \caption[Folded]{{\it Top:} Synthesized star with a flare (white circle) occurring  during the transit (planet represented by a dark circle) but at a different latitude. {\it Bottom:} Simulated planetary transit lightcurve with a significant peak due to the presence of the flare.}
\label{fig:figura1_3}
\end{figure}

There is also another situation for the detection of flares during transits. When a flare occurs in a different stellar latitude than that of the projected transit, a very significant peak will be visible in the transit. This scenario is shown in Fig. \ref{fig:figura1_3}, where the flare is located at latitude 30$^\circ$, which is different from that of the transit band (-3.7$^\circ$). As a consequence, the transit lightcurve in Fig. \ref{fig:figura1_3} shows a significant peak due to the brightening of the flare in the host star disk.

In this paper we report the detection of flares in different transits of the exoplanet Kepler-96b. These transits show only
the second situation described above: the occurrence of flares outside the occulted transit band. These flares can be characterised by a Gaussian profile described by three parameters: maximum time of the flare in the transit, flare duration, and amplitude (in units of the disk centre intensity of the star, I$_{\rm c}$). 

Our aim is to characterise the short duration high amplitude stellar flares in the transits lightcurve of Kepler-96. From the 86 transits detected, three of them (30th, 48th, and 67th) show flare signatures, and in particular transit 48th has a high intensity flare that causes an increase of $4\%$ in the total luminosity flux of the star. We have modelled the flare using the transit model described in \cite{sil03}, in which a stellar flare is added to the disk of the star outside the planet transit band. A gaussian profile was chosen to model the flare in time, $t$, as described below:

\begin{equation}
I_{\rm flare} = A \exp\left[\frac{(t - t_{0})^{2}}{2\sigma_{t}^2}\right]
\end{equation}

\noindent where $A$ is the amplitude of the flare, $t_{0}$ represents the peak time as measured from midtransit time and $\sigma_{t}$ is the temporal duration of the flare. 

\begin{figure}[!ht]
  \centering
\includegraphics[scale=0.30]{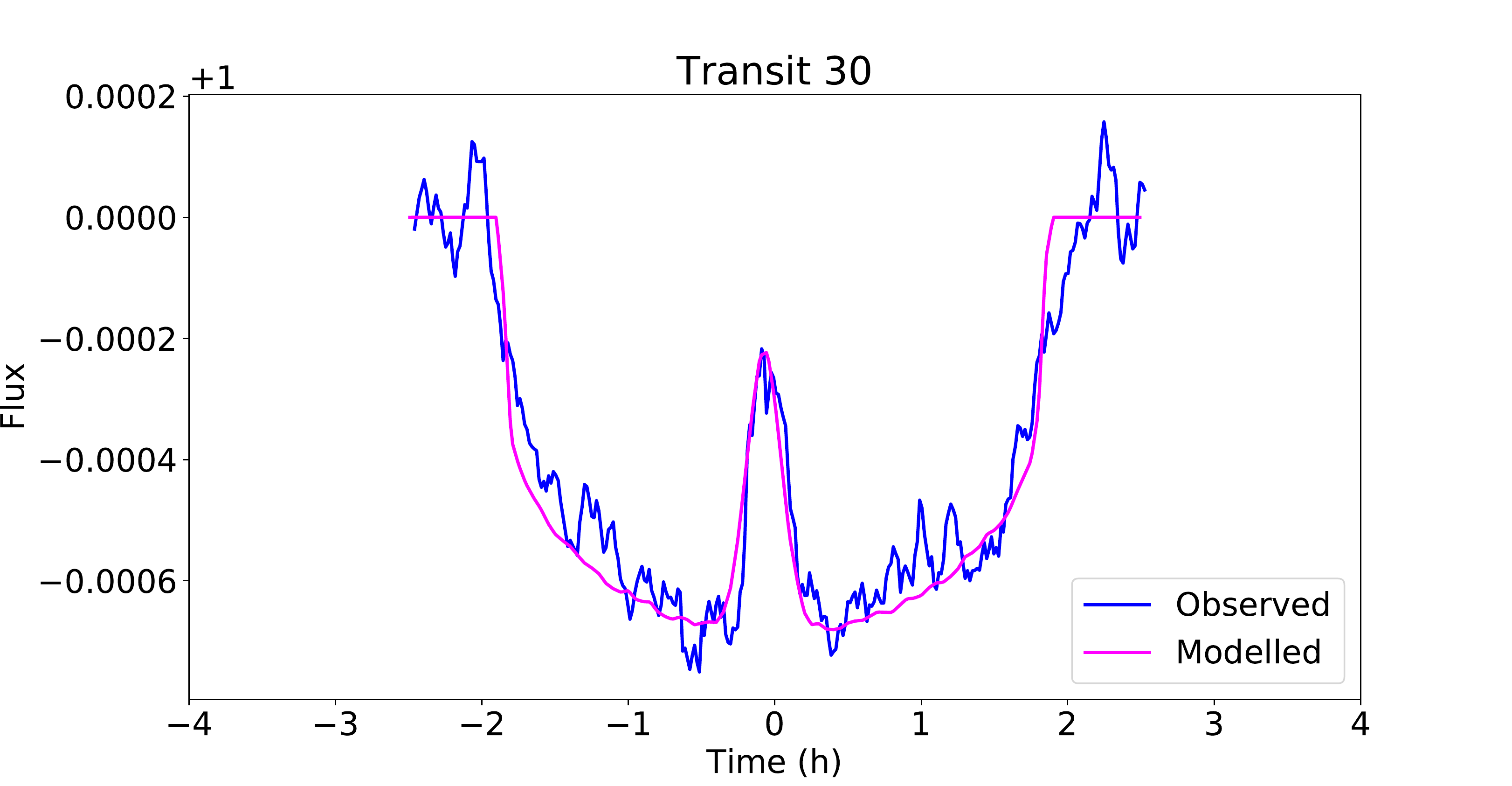} \\
\includegraphics[scale=0.30]{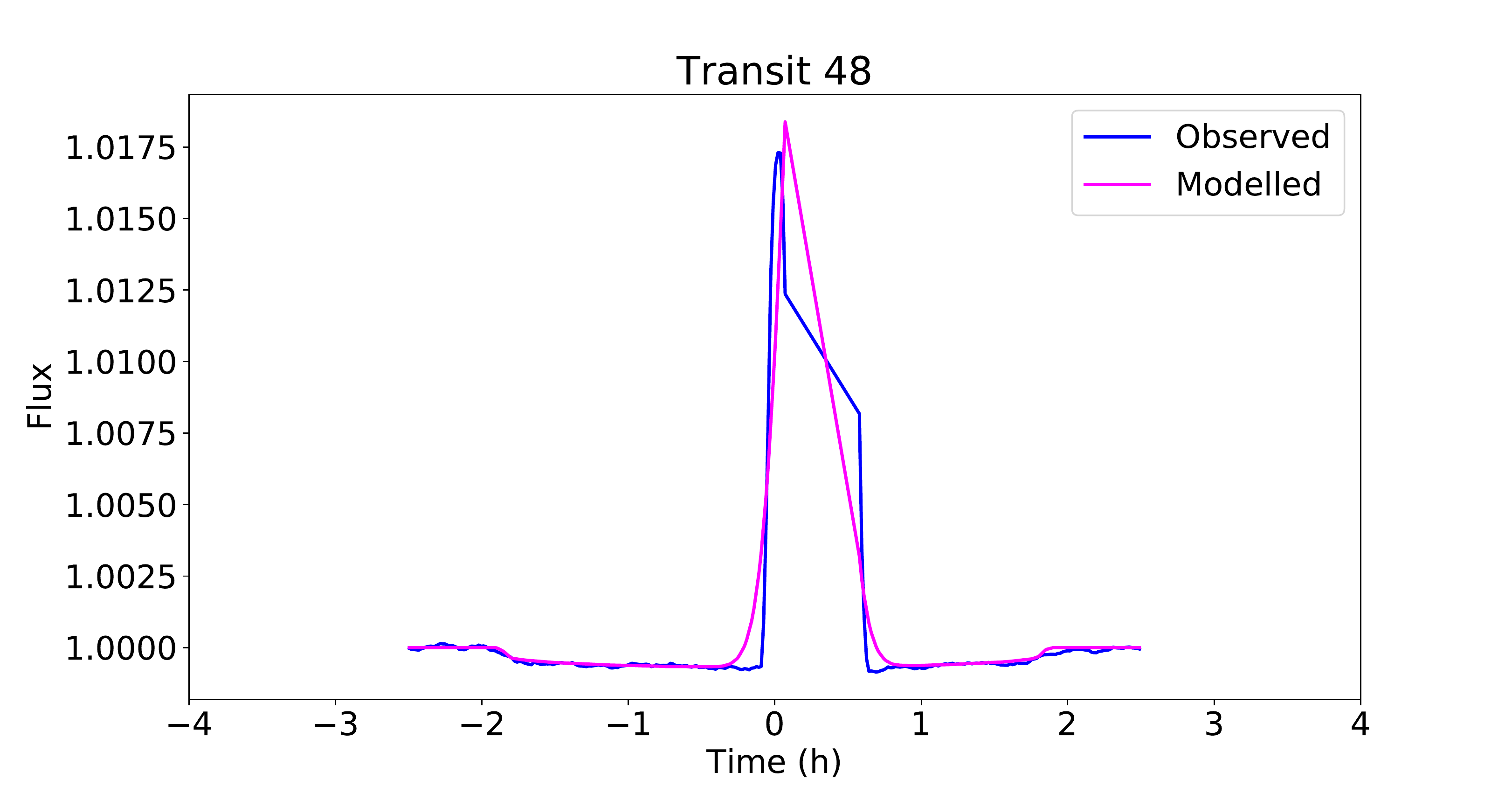} \\
\includegraphics[scale=0.30]{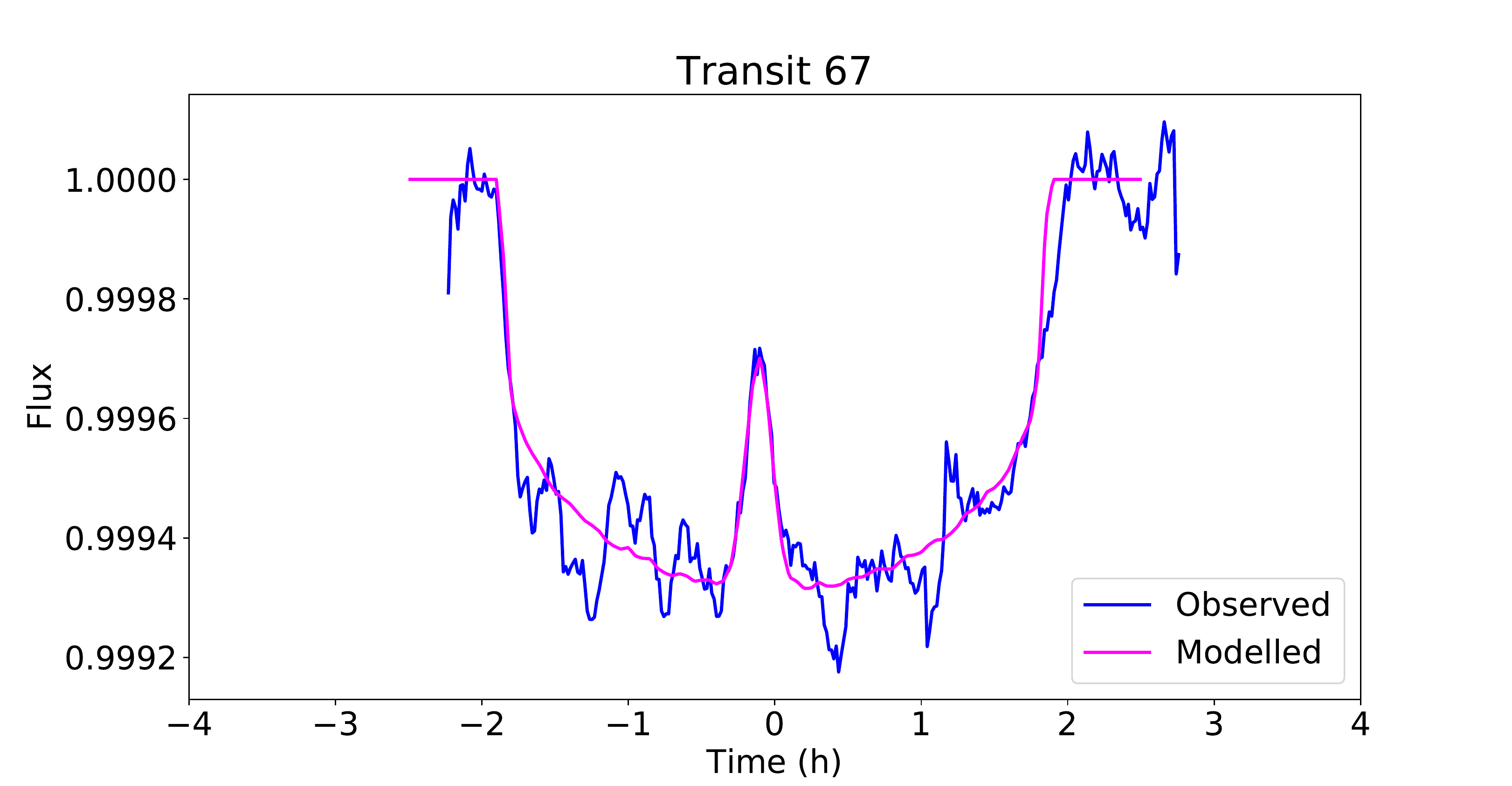} \\ 
  \caption[Folded]{From top to bottom: transits 30th, 48th, and 67th of Kepler-96b showing the increase in intensity due to the superflares (blue lines). Overplotted as pink lines are the best fit of a model of a Gaussian temporal profile stellar flare.}
\label{fig:figura1_4}
\end{figure}

The three parameters that characterise the flare ($A$, $t_{0}$, and $\sigma_{t}$) were obtained by minimising the $\chi^{2}$ between the modelled transit lightcurve and the observed data. We considered initial guesses of 0h (centered at midtransit) for $t_{0}$, 0.02h for $\sigma_{t}$ and 40.000$I_{\rm c}$ for $A$. The results are summarised in Table \ref{table:tab3}. The observed data and the fitted transit lightcurve for the three flares analysed are shown in Figure \ref{fig:figura1_4}, whereas the fit parameters are listed on Table~\ref{table:tab3}.

In addition, we computed the total energy released by the flare using the following formula:

\begin{equation}
E_{\rm flare} = A L_{\rm star} \int^{+\infty}_{-\infty} \exp(t - t0^{2}/2 \sigma_{t}^{2}) \rm dt
\end{equation}

\subsection{Results}\label{sec:32}

By modelling the flares using the transit model from \cite{sil03} (described in \ref{sec:sec31}, we found an energy release of E = 2 $\times$ 10$^{33}$ ergs for the flare in transit 30th, E = 1.8 $\times$ 10$^{35}$ ergs for transit 48th, and E = 1.2 $\times$ 10$^{33}$ ergs for transit 67th. These values are within the energy range found for superflares ($\sim$ 10$^{33}$-10$^{38}$ erg) \citep{karoff16}. Table~\ref{table:tab3} shows the values found for the peak time, time duration, and the amplitude of the three modelled superflares. 


\section{Biological Impact}\label{bio}

The total UV flux from the host star plays an important factor in the circumstances in which life originates on an orbiting planet. Stellar flares can have a significant contribution to the total UV flux of the star. For this reason, we analysed here the possible impact that the superflares produced by Kepler-96 could have in the development of a planet on its habitable zone (1 AU). 

The total photon energy in the UV band that falls in an unit area of a biological body is given by:

\begin{equation}
I_{r} = \int_{0}^{T} \int_{\lambda_{1}}^{\lambda_{2}} I(\lambda) d\lambda \,dt
\end{equation}

\noindent where $\lambda$ is the wavelength in the UV range, and $T$ is the total duration of the UV exposure.
However, the response of a biological body varies as function of the wavelength. Therefore, it is necessary to weight the incident flux by the action spectra, a function that express the biological response effectiveness at different wavelengths. 

\begin{figure}[!ht]
  \centering
\includegraphics[scale=0.70]{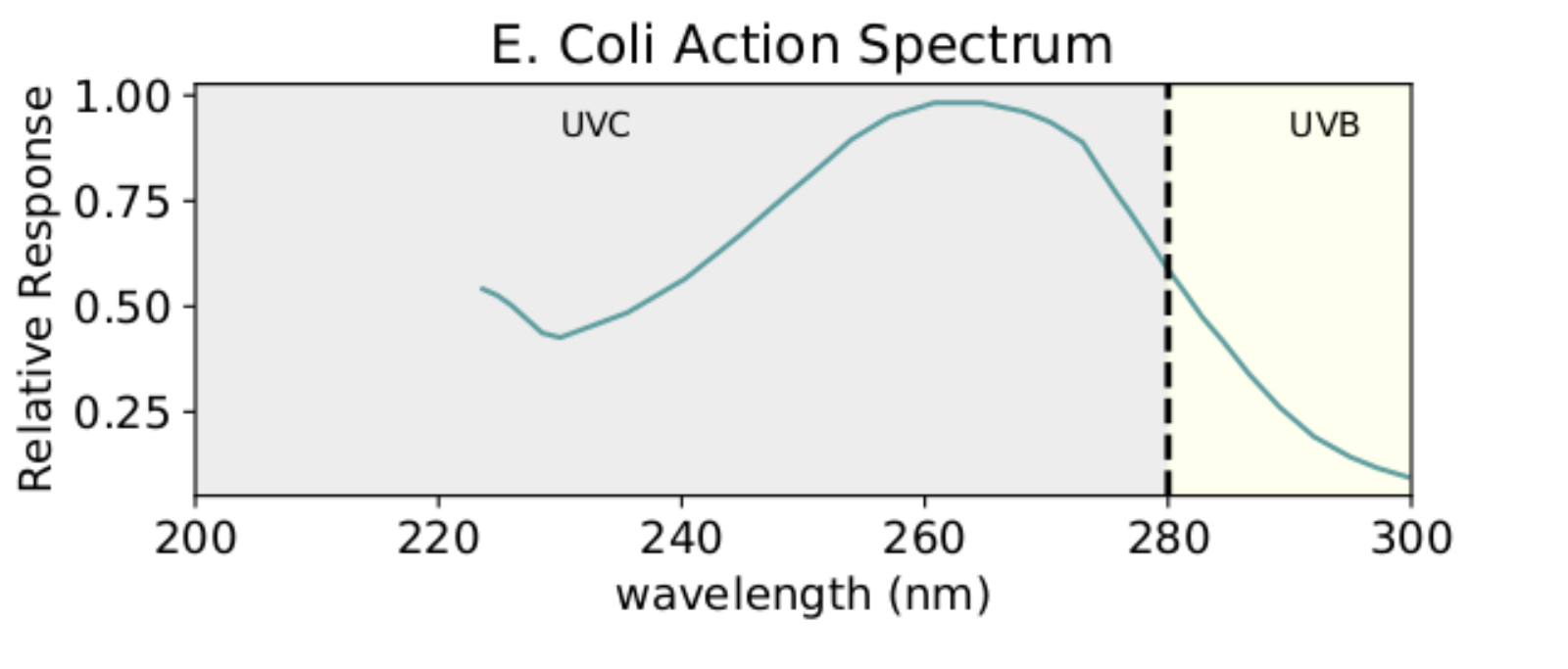} 
\includegraphics[scale=0.70]{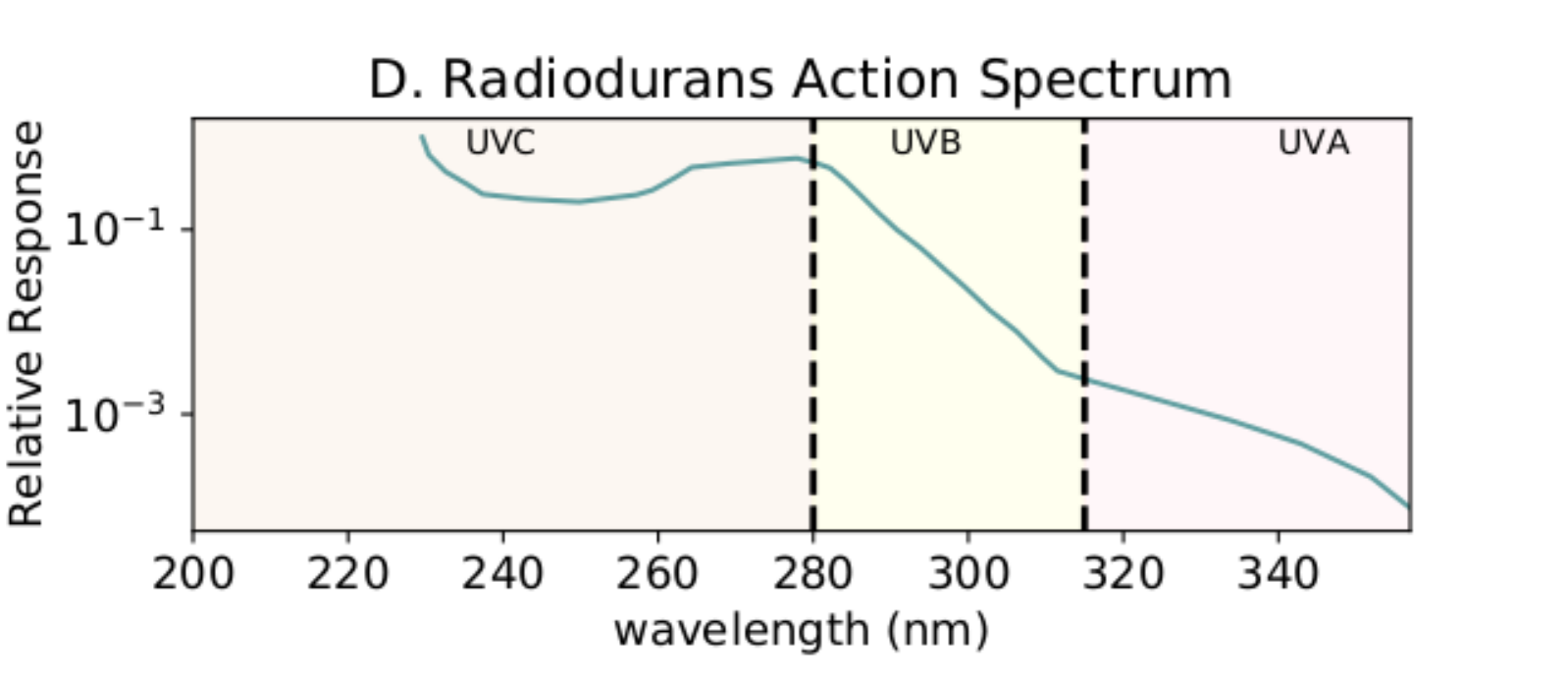} 
  \caption[Folded]{{\it Top}: {\it E.Coli} Action Spectrum for the wavelength range of 200-300nm based on \cite{giller00}. {\it Bottom}: {\it D. Radiodurans} Action Spectrum for the wavelength range of 200-300nm \citep{setlow65, calkins82,malley17}. Both action spectra are normalized at 260nm, and are higher for UVC but decreases considerably for UVB.}
\label{fig:figura1_5}
\end{figure}

Here we used the action spectrum of two microorganisms that define the surviving zone for life: {\it Deinococcus radiodurans} (\cite{setlow65, calkins82,malley17}) and {\it Escherichia coli} (\cite{giller00}) (see Fig.~\ref{fig:figura1_5}). The first one is an extremophile, one of the most resistant organisms to radiation that can survive in extreme environments conditions like vacuum, dehydration, and high dosages of ultraviolet radiation. The second one is also resistant to radiation, but less than {\it Deinococcus radiodurans}. Nevertheless, {\it E. Coli} has evolved to resist high pressures such as 2 GPa \citep{vanlint11}, and it has a genetic similarity with deep-sea bacterias \citep{horikoshi12}. Thus, we estimated the biologically effective irradiance (E$_{\rm eff}$), or fluence, if it is in J/m$^{2}$, by using the following expression:


\begin{equation}
E_{\rm eff} = \int_{\lambda_{1}}^{\lambda_{2}} F_{inc} (\lambda) S(\lambda) \,d\lambda
\end{equation}

\noindent where $F_{\rm inc}$ is the total incident UV flux including the superflare contribution arriving at the planet surface, $S$ is the action spectra, and $\lambda$ represent the middle-UV wavelengths (MUV, 200-300nm). 

For understanding the impact of these superflares on life, we are interested mainly on the MUV region of the spectra because shorter wavelengths (0.1-200nm) are attenuated in the top of the atmosphere, considering an atmosphere with strong absorbers like N$_{2}$, CO$_{2}$ or O$_{2}$. On the other hand, radiation at wavelengths within 200-320nm can partially reach the surface of the planet depending wether the planet has or not an ozone layer. This wavelength band corresponds to the UVC and UVB range which are very harmful to life. Longer wavelengths radiation can pass through the atmosphere but  are not dangerous to life. For this reason, we are going to foccus only in the superflare contribution that comes from the MUV region of the spectra, which includes the UVC and UVB.

\subsection{UV  contribution from superflares}

To estimate the UV flux contribution from Kepler-96 superflares, we used the MUV flux measured on the most intense solar flare detected, a X17 GOES class observed in 2003, with a total energy of E = 4 $\times$ 10$^{32}$ ergs. From the analysis of this flare, it was found that the contribution of the VUV (0-200nm) to the total solar irradiance was 23$\%$ \citep{woods04}. In addition, \cite{kretz11}  has also analysed the same event and claimed that the visible and  near UV fluxes (300-400nm) in the flare spectrum are the dominant energy contributions, with the visible being responsible for 70$\%$ of the total irradiated energy. This mean that the contribution to the total solar irradiation coming from the MUV (200-300nm) is around $\sim$10$\%$. Indeed, \cite{woods04} reported that this solar flare increased by 12$\%$ the Mg II h and k emissions (279.58-279.70nm), which is within the MUV range.

Therefore, we used the increase of 12$\%$ in the MUV flux measured for the most intense solar flare observed, as a proxy for the superflare MUV flux. As the total thermal blackbody flux of Kepler-96 and the Sun are very similar, we considered that the superflares found in Kepler-96 would increase the MUV flux proportionally. Therefore, a superflare found in Kepler-96 with  E = 1.8 $\times$ 10$^{35}$ ergs would increase by about 5400$\%$ the solar MUV flux (60$\%$ for E = 2 $\times$ 10$^{33}$ ergs and 36$\%$ for E = 1.2 $\times$ 10$^{33}$ ergs).

\subsection{Primitive Earth atmosphere}

The UV flux arriving at the surface of a planet is attenuated by the presence of an atmosphere. Following \cite{cnossen07}, we considered the UV surface irradiance (F$_{\rm inc}$) attenuated by atmospheres found at different Earth's epochs: an Archean atmosphere composed by 80$\%$ N$_{2}$ and 20$\%$ CO$_{2}$, with surface concentrations of 2.09 $\times$ 10$^{19}$cm$^{-3}$ and 5.24 $\times$ 10$^{18}$cm$^{-3}$ respectively, and a Present day atmosphere with ozone with concentrations of 2.09 $\times$ 10$^{19}$cm$^{-3}$ for N$_2$, 5.62 $\times$ 10$^{18}$cm$^{-3}$ for O$_2$, and 1.35 $\times$ 10$^{12}$cm$^{-3}$ for O$_3$. \cite{cnossen07} considered that the non-flaring UV irradiance at Earth's surface between 4-3.5 Gyrs ago was 75$\%$ of the present-day solar irradiation.

The total flux as a function of MUV wavelength arriving at the surface of the planet is depicted in Figure~\ref{fig:figura1_6} for an Archean (green) and Present day (purple) atmospheres. As can be seen from the plots, a Present day atmosphere absorbs basically all the radiation shortwards of 280 nm, whereas for the Archean atmosphere this is the case only for wavelengths smaller than 200 nm.
In Section~\ref{sec:surface}, we analysed the UV flux received by biological bodies (estimated by E$_{eff}$) considering the attenuation of the increased flare flux by each atmospheric scenario.

\subsection{UV Irradiation in the ocean}

During the Archean era (3.9 to 2.5 Gyr ago), the ozone layer on Earth was being formed and the UV flux arriving at the surface was probably higher than that of the present-day due to flaring activity of the younger Sun. This exposure to UV radiation could have imposed difficulties to micro-organisms survival. However, the deep ocean might have provided a safe refuge against the UV radiation. In the case of Kepler-96 system which receives much stronger UV radiation due to the superflares from the host star, the effects of the this radiation could have been attenuated for micro-organisms living in the ocean. For this reason, depending on the absorption of the UV radiation by the water, the aquatic environment is more likely to host life in an Earth-like planet orbiting Kepler-96. Therefore, we assume here that a hypothetical Earth at 1 AU orbiting the star Kepler-96 has a calm and flat Archean ocean where life could be protected.

\begin{figure}[!ht]
  \centering
\includegraphics[scale=0.40]{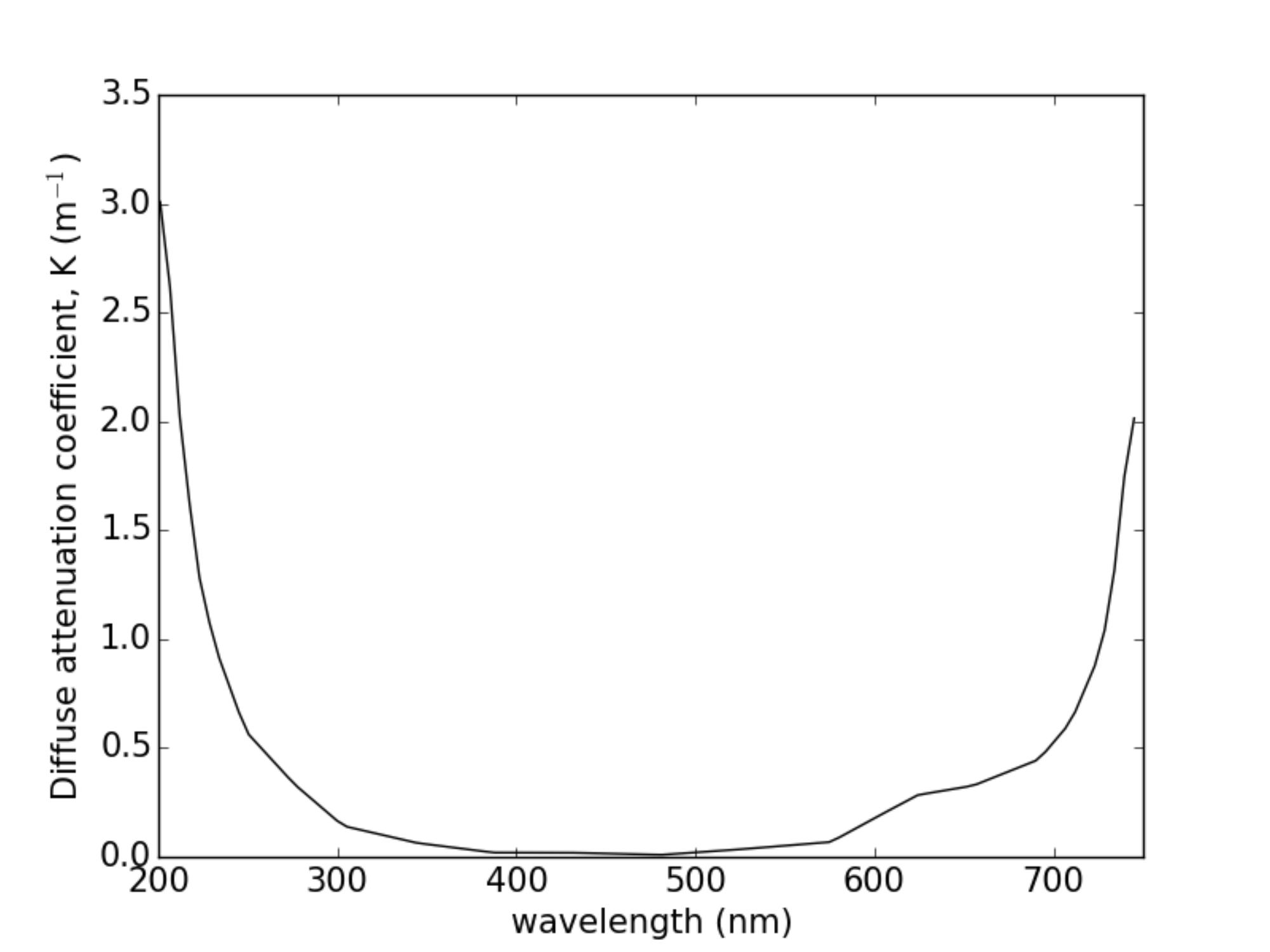} 
  \caption[Folded]{Diffuse attenuation coefficients from \cite{smith81} for the clearest natural water from 200 to 750nm used to estimate the UV radiation penetration in the Archean ocean.}
\label{fig:figura1_7}
\end{figure}

The UV radiation is absorbed and scattered when propagating through the water. The resulting flux at depth $z$ is given by:

\begin{equation}
I(\lambda, z) = I_{0}(\lambda) e^{-K(\lambda),z}
\end{equation}

\noindent where $I(\lambda, z)$ is the UV spectral irradiance at depth $z$, $I_{0}(\lambda)$ is the UV spectral irradiance with the superflare contribution at the ocean surface that has passed through an Archean atmosphere, and $K(\lambda)$ is the diffuse attenuation coefficient for water given by the sum of the absorption coefficient of water and the scattering coefficient \citep{kirk94, cockell00}. The diffuse attenuation coefficient determines the decay rate of the UV radiation with depth. In this paper we used the values determined by \cite{smith81} for the diffuse attenuation coefficient, $K(\lambda)$, computed for the clearest natural water in the spectral region from 200 to 750nm. These values are shown in Figure \ref{fig:figura1_7}. 

To estimate at which ocean depth {\it E. Coli} and {\it D. Radiodurans} could tolerate the UV dosage received by them, we computed the E$_{\rm eff}$ by convolving the attenuated irradiation by the water (I$_{0}$) with the action spectrum from these microorganisms, shown in Figure~\ref{fig:figura1_5}.


\section{Results and discussion}
\label{sec:sec4}

\subsection{On the surface}\label{sec:surface}

\begin{figure*}[!ht]
	\begin{center}
        \subfigure[]{%
            \label{fig:first}
            \includegraphics[width=0.70\textwidth]{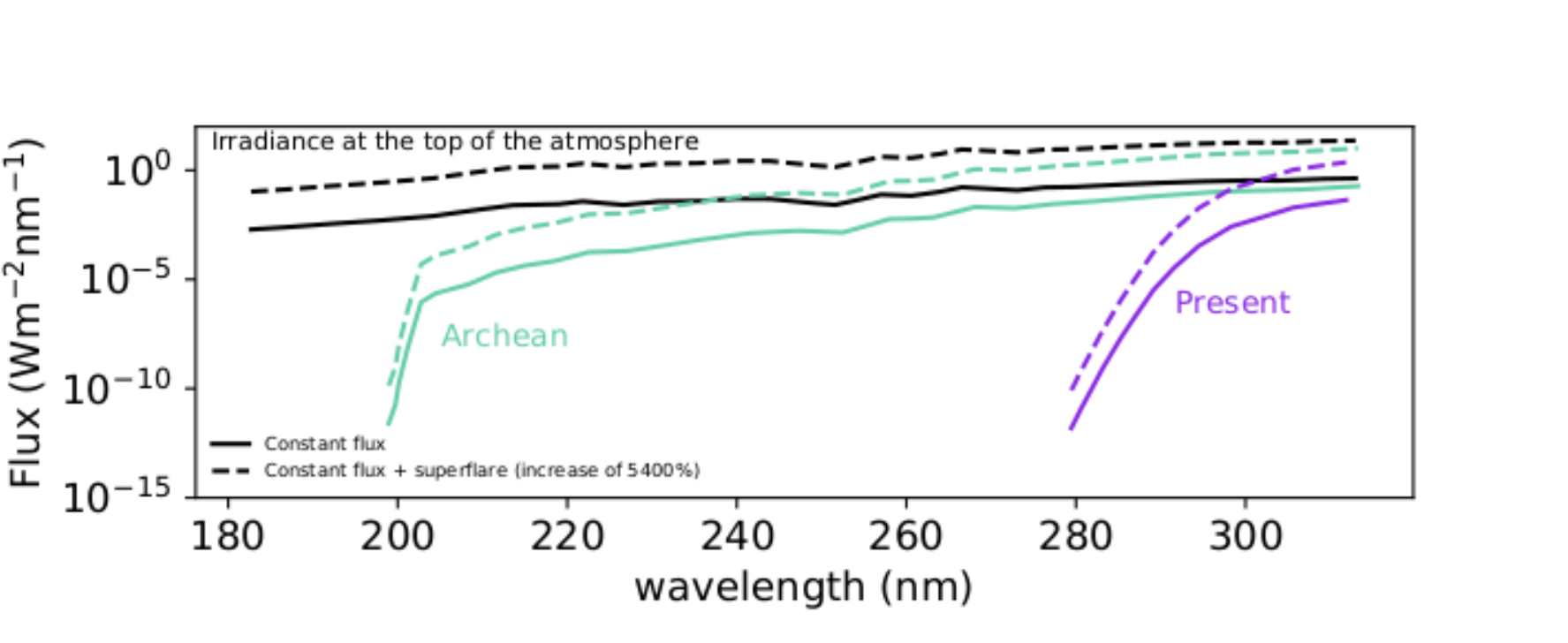}
        }\\%
        \subfigure[]{%
           \label{fig:second}
           \includegraphics[width=0.70\textwidth]{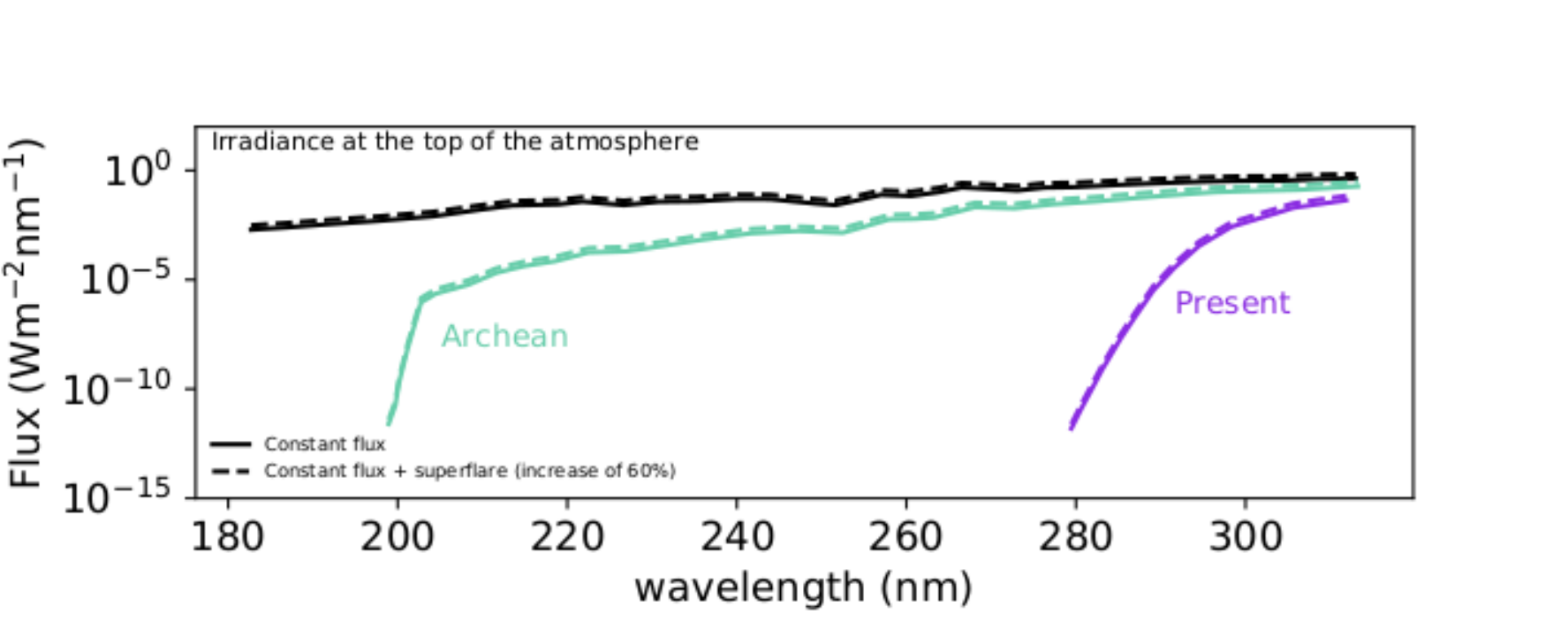}
        }\\ 
     \end{center}
  \caption[Folded]{Surface Irradiance after being attenuated by different atmospheres found on Earth: an Archean atmosphere (green) and a present day atmosphere with ozone (purple). The contribution of the superflare for the UV flux is indicated by a dashed line and the rescaled UV flux (with the contribution of the superflare) arriving at the surface of Kepler-96b  is represented by a dotted line. From top to bottom: for a superflare UV flux contribution of (a) 5400$\%$ (flare B), (b) 60$\%$ (flare A). }
\label{fig:figura1_6}
\end{figure*}

We analysed the increased MUV flux being attenuated by a planet in the habitable zone with an Archean atmosphere or a Present-day atmosphere with ozone, adopted from \citep{cnossen07}. These scenarios are shown in Figure \ref{fig:figura1_6} (a, b for contributions from superflares A and B), where the increased UV flux by superflares is represented by a dashed line. It is important to note that for an Archean atmosphere there is still flux in the UVC band which can be harmful to life.

To analyse the effects of the attenuated radiation, we estimated the biologically effective irradiance, E$_{\rm eff}$ (explained in detail in Section \ref{bio}) using the action spectra of {\it Deinococcus Radiodurans} and {\it Escherichia Coli} (see Fig.~\ref{fig:figura1_6}), which gives the overall effectiveness of the UV flux in a biological body.
The threshold for  E$_{\rm eff}$ was chosen using the maximum UV flux for 10$\%$ survival of these bacteria. These values are  $F_{10}^{\rm UV} = 553$ J/m$^{2}$ for {\it D. Radiodurans} \citep{ghosal05}, and for {\it E. Coli}, less resistant to radiation,  $F_{10}^{\rm UV}$= 22.6 J/m$^{2}$ \citep{gascon95}. 

Table \ref{table:tab4} summarizes the results found for the biological effectiveness fluence for these bacterias considering the UV contributions of the superflares analysed in this work. Thus, from the results in Table \ref{table:tab4} it is possible to see that for a sudden increase of 5400$\%$ (from flare B) in the incident UV flux, the only chance for both micro-organisms to survive on the planet surface at 1AU was if its atmosphere had an ozone layer. For an increase of 60$\%$ or 35$\%$ (from flares A and C), lifeforms such as {\it D. Radiodurans} could thrive even when the planet had an Archean atmosphere.

\subsection{In the ocean}

To be suitable for life, a certain depth of the ocean should not receive considerable UV flux, in particular UVB. The exposure to UVB can be very dangerous for life, so only very low values of this radiation should reach the biological body. In the case of cyanobacteria, the first photosynthetic oxygen-evolving organisms, UVB can cause harmful effects, in particular on its DNA and photosynthesis \citep{singh10}. 

\begin{figure}[ht]
  \centering
            \includegraphics[width=0.72\textwidth]{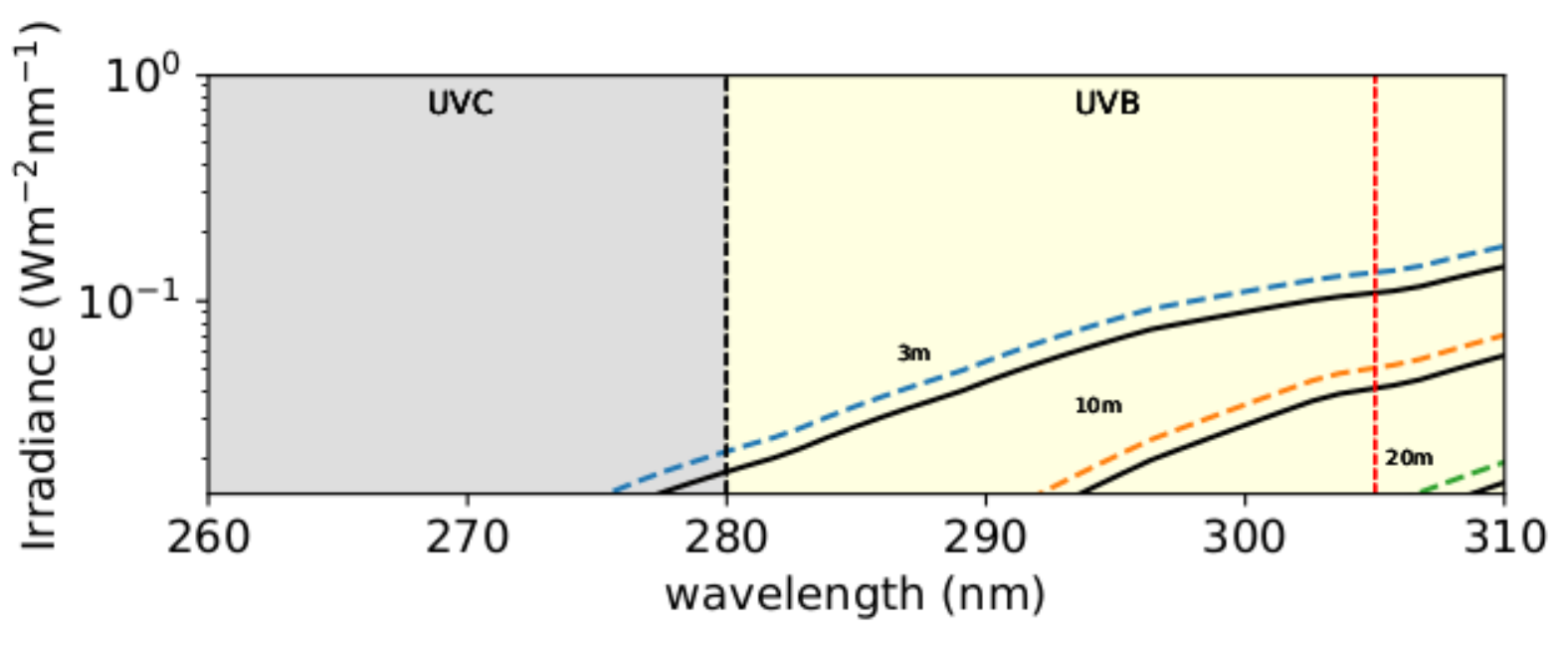}
            \includegraphics[width=0.72\textwidth]{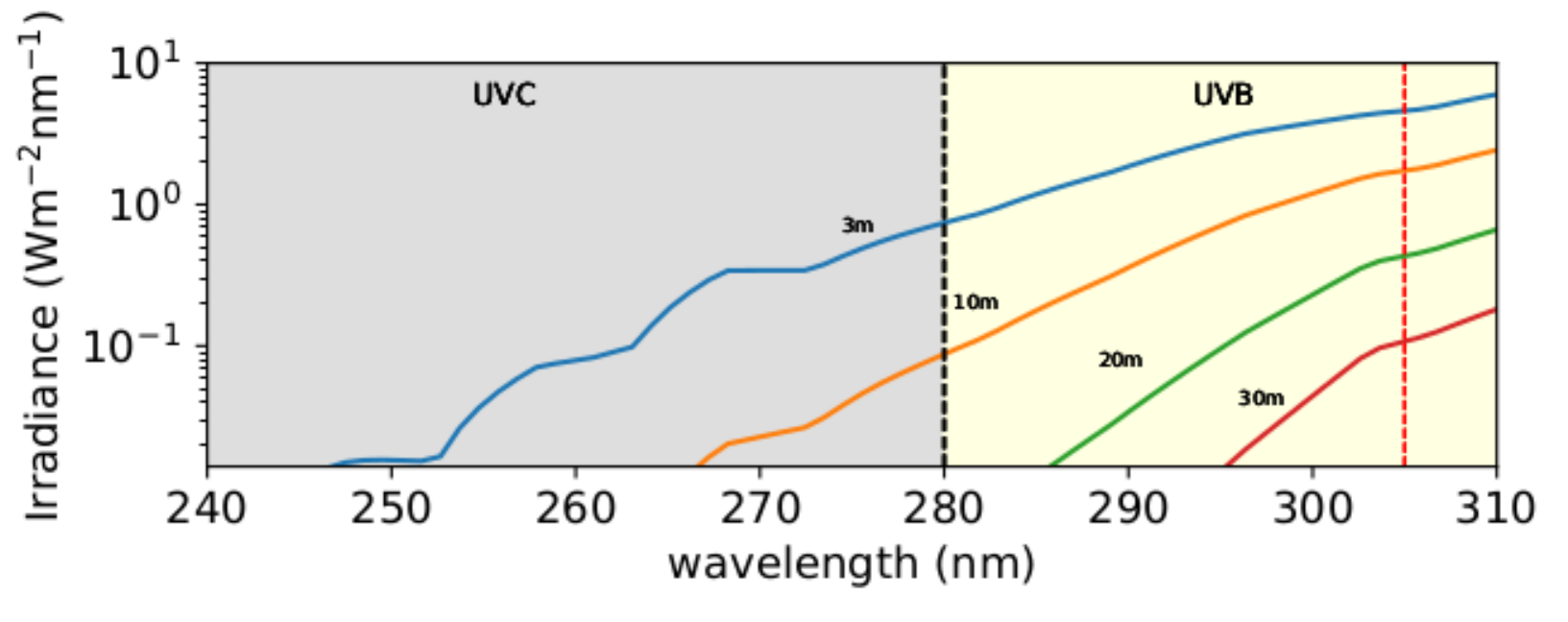}
  \caption[Folded]{Kepler-96 UV flux irradiation after the superflare contribution reaching various depths of an Archean ocean in a hypothetical Earth in the HZ of Kepler-96. {\it Top}: for superflares of E = 1.2 $\times$ 10$^{33}$ ergs (solid lines) and E = 2.0 $\times$ 10$^{33}$ ergs (dashed lines). {\it Bottom}: considering the UV flux increase by the strongest superflare (E = 1.8 $\times$ 10$^{35}$ ergs) . The same UVB and UVC irradiation that we find on the present-day surface of the Earth's ocean at 305nm (indicated by a red dashed vertical line) is present at a depth of 3m (top) and 30m (bottom) of this ocean.}
\label{fig:figura1_8}
\end{figure}

The values of UV radiation varying with ocean depths are shown in Figure \ref{fig:figura1_8}. The UVB irradiance found at the present-day ocean surface on Earth is $\sim$ 0.1 (at 305 nm) Wm$^{-2}$nm$^{-1}$ \citep{smyth11}. In the presence of the superflares, the depths at which we find similar values of UVB is 3m deep for the hypothetical Earth analysed here (see Fig.~\ref{fig:figura1_8}b), considering an increase of 35$\%$ or 60$\%$ of the UV flux, or 30m deep, if the increase is of 5400$\%$ (see Fig.~\ref{fig:figura1_8}a). 

\begin{figure}[!ht]
  \centering
\includegraphics[scale=0.85]{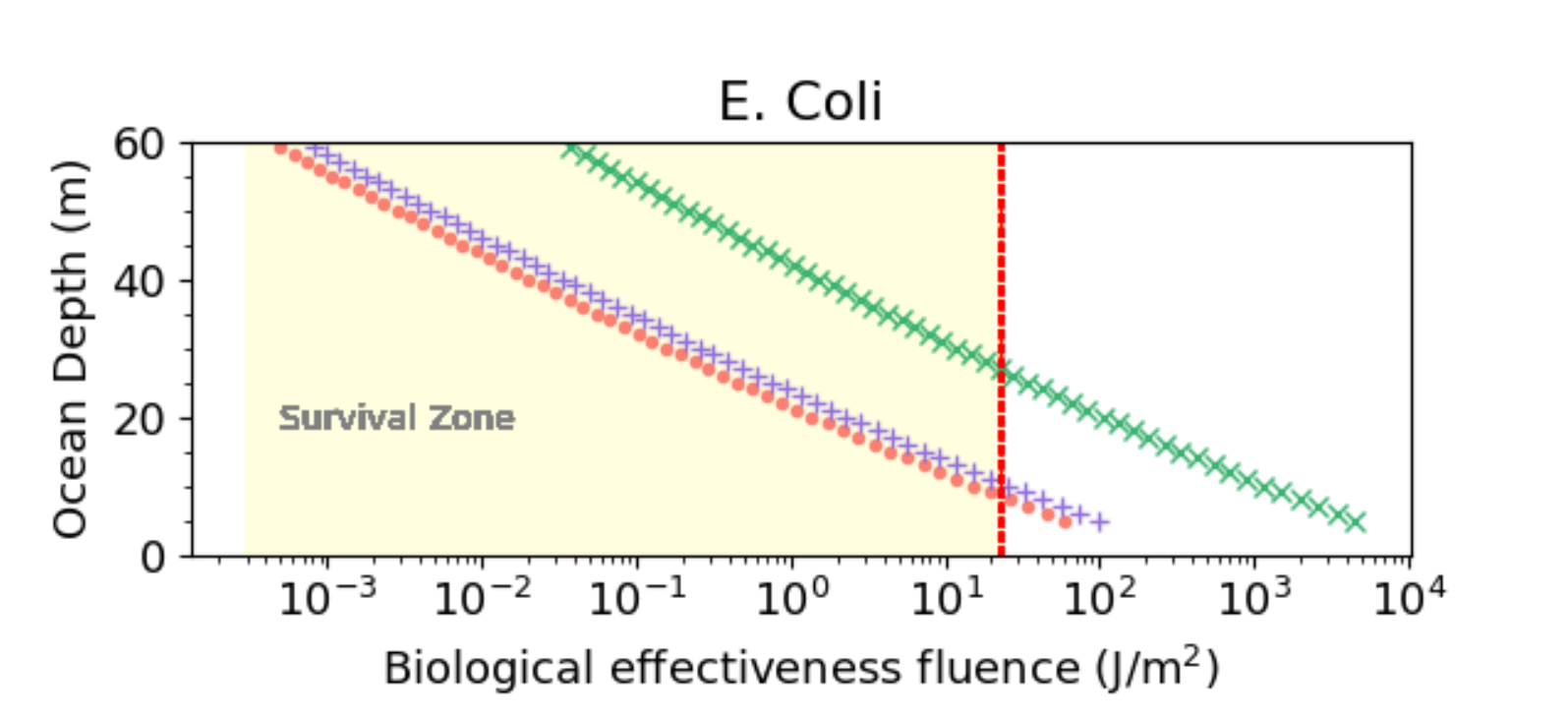}
\includegraphics[scale=0.85]{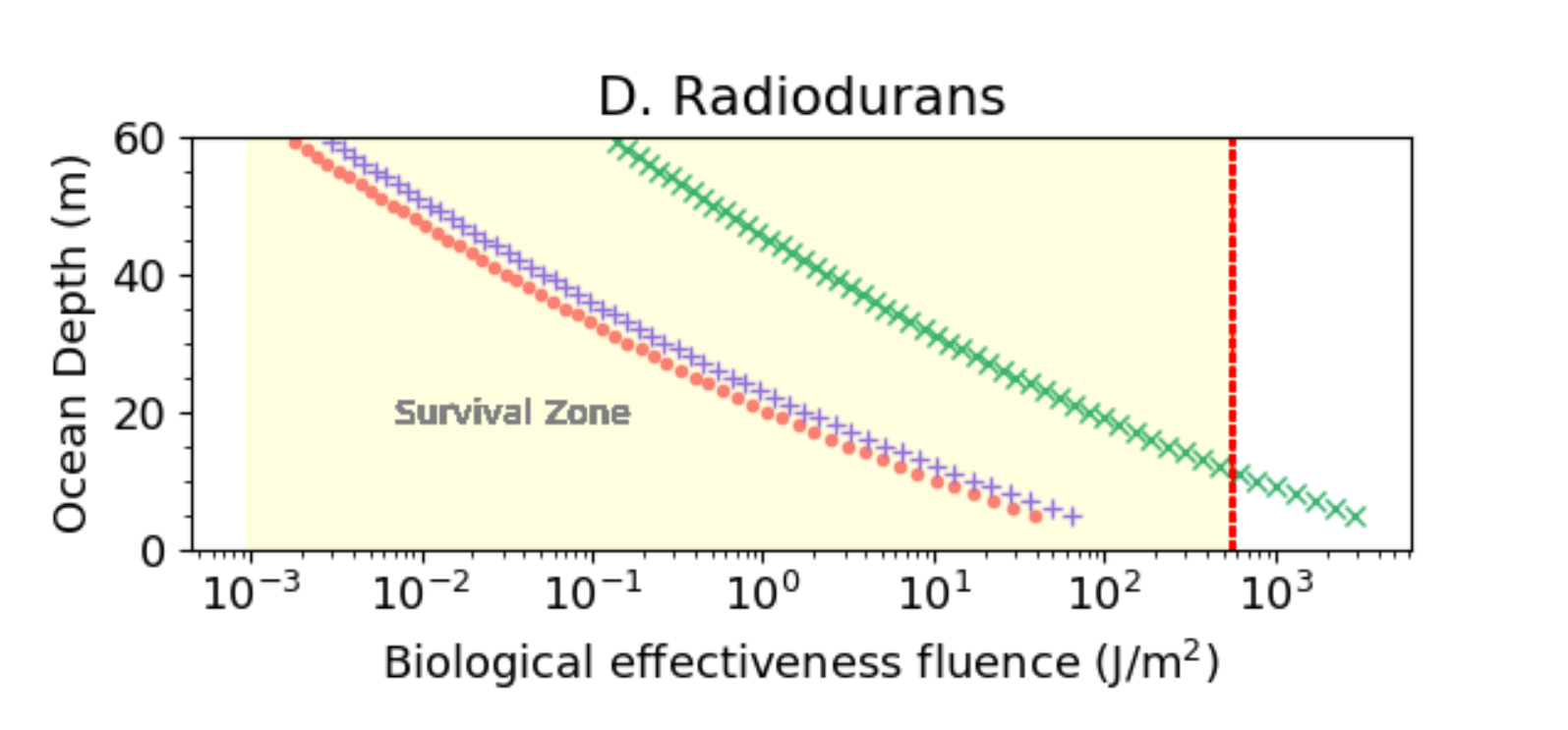}
  \caption[Folded]{{\it Top:} Biologically effective irradiance for {\it E. Coli} (E$_{\rm eff}$) with depth in an Archean ocean present in a planet in the HZ of Kepler-96. The values in Joules of the E$_{\rm eff}$ were obtained by multiplying the values in Watts with the time duration of the superflares, which are 7.1 min (flare A), 9.7 min (flare B) and 5.3 min (flare C). The E$_{\rm eff}$ computed with the contribution of each superflare is represented as follows: pink dot (flare C), purple cross (flare A), green x symbol (flare B). {\it Bottom}: the same for {\it D. Radiodurans}. The red vertical line represents the threshold for finding life determined by the maximum UV flux for 10$\%$ survival of {\it E. Coli} and of {\it D. Radiodurans}.}
\label{fig:figura1_9}
\end{figure}

To verify the UV impact directly on {\it E. Coli} and {\it D. Radiodurans}, we weighted the increased UV flux attenuated by water (I$_{0}$) with their action spectrum. Then, to check if they can survive the amount of UV that they are receiving, we used as threshold the maximum UV flux for 10$\%$ survival of {\it E. Coli} $F_{10}^{\rm UV}$= 22.6 J/m$^{2}$ and for {\it D. Radiodurans} $F_{10}^{\rm UV}$= 553 J/m$^{2}$ to determine the ocean depth for which these microorganisms would not be damaged. Despite the fact {\it D. Radiodurans} is not found in the ocean, \cite{cockell00} proposed that Archean cyanobacteria could have been UV-resistant with tolerances to the UV similar to those of the {\it D. radiodurans}. This result is shown in Figure \ref{fig:figura1_9} for an ocean in the planet in the HZ, where the threshold is represented by a red vertical line.

Thus, considering the effects of the strongest superflare (flare B) in a planet in the HZ, an ocean depth of 28m would be necessary to absorb the UV irradiation such that {\it E. Coli} could survive and 12m for {\it D. Radiodurans}. However, if we take into consideration the effects of a superflare of flare C or flare A, lifeforms such as {\it E. Coli} would be possible at lower depths of around 10m, and {\it D. Radiodurans} could live on the planet surface. All the ocean depths found here are within the photic zone (200m deep), the top layer of the ocean on Earth that receives sufficient luminosity to permit photosynthesis and is where most of the marine life lives. On Earth's ocean, is also possible to find life in the aphotic zone (200-1000m under the ocean surface), where life is adapted to the darkness, cold and high pressure. 

\subsection{Frequency of the superflares}

Flares can be associated with the release of high energetic protons. If these protons arrive on the planet, it can cause the production of HO$_{x}$ and NO$_{x}$ that leads to ozone depletion \citep{rohen05}. Fortunately, not all flares events will result in a energetic proton event. The probability of such proton events is dependent on the flare energy, therefore they will be more likely to occur on flares with energy of 10$^{28.3}$ ergs and above \citep{yashiro06, hudson11, dierc15, tilley17}

In Segura (2010) is shown that the ozone layer recovers in about 30 hrs if there are no energetic particles associated with the flare. If there are energetic particles ejected during the flare then the ozone layer is almost totally depleted for about 30 years. 

Both time intervals for ozone recovering would be sufficient when we take into account the average occurrence frequency of superflares on Sun-like stars. From the statistical analysis of 187 flares on 23 solar-type stars, \cite{maehara15} found that flares with bolometric energy of 10$^{33}$, 10$^{34}$, and 10$^{35}$ erg occur once in approximately 70 years, approximately 500 years, and approximately 4000 years, respectively.

However, if we look to the time interval between the superflares found in this work (inside transits), which is $\sim$ 300 days, we have a sufficient time interval for the ozone be recovered in the case of superflares not associated with particle ejections. If we take into consideration only superflares with E $>$ 10$^{33}$ localized within or outside the transit, one can observe 3 of them in the lightcurve of Kepler-96. The time interval between the first two is about 30 days, while the second and the last has 338 days between them. Again, both time intervals would be sufficient for the ozone layer recover only for superflares not associated with particle ejections.

\section{Conclusions}
\label{sec:sec5}

In this work, we detected and characterized stellar superflares present in transit lightcurves and analysed their impacts on extremophile lifeforms. The star analysed here, Kepler-96, is a solar-type star ($M = 1.00 \pm 0.06M_{\odot}$ and $R = 1.02 \pm 0.09R_{\odot}$) with rotation period of $P_{\rm rot}$ = 11.89 days and an age of 2.4 Gyr which is the same as the Sun when oxygen appeared on Earth's atmosphere. Moreover, this star is accompanied by a transiting super-Earth planet in close orbit: Kepler-96b. Analysing the transits of Kepler-96b, we detected superflares in 3 of them. We modelled these superflares using the transit model from \cite{sil03}, which allows the insertion of flares in the stellar disc. Therefore, we estimated their physical characteristics (time of maximum, time duration, and amplitude), whereas the strongest superflare yield an energy release of E =  1.8 $\times$ 10$^{35}$ ergs. 

In addition, we estimated the biological impacts that the increased UV flux coming from these superflares could have in a hypothetical Earth in the habitable zone of Kepler-96 either for life on the planet surface or in the ocean. In both scenarios, we estimated the  biological effectiveness of the UV flux (E$_{\rm eff}$) in two resistant microorganisms, {\it E. Coli} and {\it D. Radiodurans}, by weightening the UV flux with their action spectrum. To estimate the E$_{\rm eff}$ we also considered the increased UV flux attenuated by two atmospheric scenarios: an Archean and a Present-day atmosphere. 

For a superflare contribution of 5400$\%$ (from flare B) in the UV flux, the biological effective irradiance shows that {\it E. Coli} and {\it D. Radiodurans} would only survive on the surface of the hypothetical Earth if there is an ozone layer present on the planet atmosphere. For superflare UV contributions of 60$\%$ and 35$\%$ (from flares A and C), {\it D. Radiodurans} could live even with an Archean atmosphere, while {\it E. Coli} would still need an ozone layer to survive. However, these superflares allied with other stellar magnetic activity phenomena (\textit{i.e}, Coronal Mass Ejections, CMEs) could be responsible for a significant depletion of the ozone layer of this planet or could even be stripping away its atmosphere. 

Indeed, the presence of a ozone layer is very important to protect life. Even if life is not possible in the surface of the planet, the ozone layer can be generated by photosynthetic organisms living in the ocean. Consequently, allowing life to migrate to the surface after some period. It is believed that this situation happened on the primitve Earth, where the development of an ozone layer occurred around 2.4 Gyrs ago (Archean era) due to the increase of oxygen in Earth's atmosphere. This saturation of oxygen was biologically induced by oceanic cyanobacteria, single-celled organisms that are believed to be the first to produce oxygen by photosynthesis. A similar scenario could be found in the Kepler-96 system. For this reason, we analysed the ocean depths that could harbour life in the hypothetical Earth. An ocean in the planet in the HZ could protect it to the increased UV radiation due to the superflares and would allow life in depths within the photic zone (up to 200m). In the case of the Early Earth, the photic zone of the Archean oceans played an important region for marine primary productivity \citep{cockell00}.

\section{Acknowledgments}

This work has been supported by grant from the Brazilian agency FAPESP ($\#$2013/10559-5) and from MackPesquisa. Raissa Estrela acknowledges a FAPESP fellowship ($\#$2016/25901-9).

\section{Author Disclosure Statement}

No competing financial interests exist.

\bibliography{bibliography}{}
\bibliographystyle{apalike}

\newpage


\begin{table*}[!ht]
\centering
\label{my-label}
\refstepcounter{table}\label{table:tab1}
\resizebox{0.70\textwidth}{!}{\begin{minipage}{\textwidth}
\begin{tabular}{cccccc}
\hline
\hline
\multicolumn{6}{c}{f{Table I}}                                                                                               \\
\multicolumn{6}{c}{Observational Parameters of the Kepler-96 star}                                                                          \\
Mass  {[}M$_{\odot}${]} & Radius {[}R$_{\odot}${]} & Age {[}Gyrs{]} & Temperature {[}K{]} & Rotational Period {[}days{]} & References \\
1 $\pm$ 0.06            & 1.02 $\pm$ 0.09       & 2.34           & 5690 $\pm$ 73       & 15.3                         & 1,2    
\end{tabular}
\begin{tablenotes}
\item \textbf{References.} (1) \cite{marcy14}, (2) \cite{walko13}.
\end{tablenotes}
\end{minipage}}
\end{table*}

\vspace{1cm}


\begin{table*}[!ht]
\label{my-label}
\refstepcounter{table}\label{table:tab2}
\resizebox{0.65\textwidth}{!}{\begin{minipage}{\textwidth}
\begin{tabular}{cccccccc}
\hline
\hline
\multicolumn{8}{c}{\textbf{Table II}}                                                                                                                                                              \\
\multicolumn{8}{c}{Observational Parameters of the Kepler-96b planet}                                                                                                                                        \\
Mass  {[}M$_{Jup}${]} & Radius {[}R$_{Jup}${]} & Radius {[}R$_{*}${]} & Orbital Period {[}days{]} & Inclination {[}deg{]} & Semi-major axis {[}AU{]} & Semi-major axis {[}R$_{star}${]} & References \\
0.027 $\pm$ 0.011     & 0.243 $\pm$ 0.001$^{*}$   & 0.024 $\pm$ 0.001$^{*}$               & 16.23                     & 90.14 $\pm$ 0.03$^{*}$              & 0.0047 $\pm$ 0.0005$^{*}$              &  34.16 $\pm$ 0.05$^{*}$                       & 1         
\end{tabular}
\begin{tablenotes}
\item $^{*}$ Values obtained in this work.
\item \textbf{References.} (1) \cite{marcy14}.
\end{tablenotes}
\end{minipage}}
\end{table*}

\vspace{1cm}


\begin{table*}[!ht]
\refstepcounter{table}\label{table:tab3}
\resizebox{0.65\textwidth}{!}{\begin{minipage}{\textwidth}
\begin{tabular}{lccccccc}
 \hline
 \hline
\multicolumn{8}{c}{\textbf{Table III}}                                                                                                                                                        
 \\
\multicolumn{8}{c}{Characteristics of the Kepler-96 superflares}                                                                                                                                            \\
Flare & Transit             & Midtransit {[}BJD -  2.454.833 days{]} & Time position {[}hours{]} & Time duration {[}hours{]} & Time duration {[}min{]} & Amplitude {[}I$_{c}${]} & Energy  {[}ergs{]}      \\
A & 30th & 674.41007                              & -0.0714 $\pm$ 0.006       & 0.118 $\pm$ 0.006         & $7.1 \pm 0.4$                     & 39627 $\pm$ 0.00002      & 2.0 $\times$ 10$^{33}$  \\
B & 48th & 966.70310                              & 0.2402 $\pm$ 0.006        & 0.161 $\pm$ 0.003         & $9.7 \pm  0.2 $                    & 2986143 $\pm$ 0.002      & 1.8 $\times$ 10$^{35}$ \\
C & 67th & 1275.2347                              & -0.108 $\pm$ 0.016        & 0.088 $\pm$ 0.017         & $5.3 \pm 1.0$                     & 32885 $\pm$ 0.00006      & 1.2 $\times$ 10$^{33}$
\end{tabular}
\end{minipage}}
\end{table*}

\vspace{1cm}


\begin{table*}[!ht]
\centering
\refstepcounter{table}\label{table:tab4}
\resizebox{0.75\textwidth}{!}{\begin{minipage}{\textwidth}
\label{table:tab4}
\begin{tabular}{lccc}
\hline
\hline
\multicolumn{4}{c}{\textbf{Table IV}}         \\                                                                                                                                                                             
\multicolumn{4}{c}{Biological effectiveness fluence from Kepler-96, E$_{\rm eff}$ {[}J/m$^{2}${]}$^\dagger$}  \\                                                                                                                           
      & No atmosphere      & Archean atmosphere & Present atmosphere with O$_{3}$ \\
      \hline
 & \multicolumn{3}{c}{Contribution of 5400$\%$ to the UV flux (Flare A)}      \\                                                                                                                                                       
E. Coli                        & 1.4 $\times$ 10$^{5}$                 & 2 $\times$ 10$^{4}$                                 &  22                                            \\
D. Radiodurans                & 8 $\times$ 10$^{4}$                   & 1.3 $\times$ 10$^{4}$                                     & 7.5                                          \\
\hline
 & \multicolumn{3}{c}{Contribution of 60$\%$ to the UV flux (Flare B)}                                                                                                                                                                 \\
E. Coli                        & 3 $\times$ 10$^{3}$                 & 455                                               & 0.5                                         \\
D. Radiodurans                 & 2 $\times$ 10$^{3}$  & 282  & 0.16                                      \\
\hline
 & \multicolumn{3}{c}{Contribution of 35$\%$ to the UV flux (Flare C)}                                                                                                                                                                 \\
E. Coli                        & 1968              &  286  & 0.3                                          \\
\multicolumn{1}{l}{D. Radiodurans} & 1127  & 177 & 0.10 
\end{tabular}
\begin{tablenotes}
\item The error values were disregarded as they are negligible.
\item $^\dagger$ To obtain the values in Joules, we multiplied the values in Watts by the total duration of the flare.
\end{tablenotes}
\end{minipage}}
\end{table*}

\newpage

\end{document}